\documentclass[12pt]{article}
\usepackage{amsmath}
\usepackage{amssymb}
\usepackage{amsfonts}
\usepackage{amscd}
\usepackage[all]{xy}

  \usepackage{makeidx}
  \usepackage{graphicx,graphics,color}
\usepackage{latexsym,doc,fontenc,exscale}
\usepackage{amsmath,amsfonts,amsthm,eucal,amssymb}

\usepackage{multicol}
\usepackage{times}
\usepackage{pstricks}
\title{Bicomplex Hamiltonian systems in Quantum Mechanics}
\author{Bijan Bagchi$^{1,}$\thanks{Electronic address: bbagchi123@gmail.com} \ and Abhijit Banerjee$^{2,}$\thanks{Electronic address: abhijit.banerjee.81@gmail.com}\\
{\small\sl $^1$ Department of Appplied Mathematics, University of Calcutta,}\\
{\small \sl 92 Acharya Prafulla Chandra Road, Kolkata 700 009, India}\\
{\small\sl $^2$ Department of Mathematics, Krishnath College,} \\
{\small \sl Berhampore, Murshidabad, India-742101}}
\date{ }
\begin{document}
{
\baselineskip=22pt plus 1pt minus 1pt
\maketitle

\begin{abstract}
We investigate bicomplex Hamiltonian systems in the framework of an analogous version of the Schr\"odinger equation. Since in such a setting three different types of conjugates of bicomplex numbers appear, each is found to define in a natural way, a separate class of time reversal operator. However, the induced parity ($\mathcal{P}$)-time ($\mathcal{T}$)-symmetric models turn out to be mutually incompatible except for two of them which could be chosen uniquely. The latter models are then explored by working within an extended phase space. Applications to the problems of harmonic oscillator, inverted oscillator and isotonic oscillator are considered and many new interesting properties are uncovered for the new types of $\mathcal{PT}$ symmetries.
\end{abstract}
\vspace{0.5cm}


\newpage
\section{\label{intro}Introduction}
\numberwithin{equation}{section}
At a fundamental level numerous extensions of quantum mechanics have been envisaged to meet the growing challenges that the theory has thrown up from time to time. Inclusion of the field of quaternions was an important advancement made by Birkhoff and von Neumann \cite{2} about eighty years ago to represent the pure states of a quantum system on any associative division algebra (see also \cite{f1,f2,AD,Mar} and the references therein). Against such a pursuit the commutative ring of bicomplex numbers \cite{3} has emerged as a viable discipline and indeed found a number of applications over the years in different directions of quantum theory \cite{4,5,6,7,8}.

With the advent of models of parity $(\mathcal{P})$-time $(\mathcal{T})$ symmetry \cite{ben1,ben2}, also claimed \cite{new} to be a plausible alternative to the requirement of Hermiticity that is implicitly relied upon as a guiding axiom in the standard quantum mechanics picture \cite{sch}, there have been not only some active theoretical developments but more importantly some experimental ones as well (\cite{n8}-\cite{n14}). Further, realization of $(\mathcal{P})$-time $(\mathcal{T})$ symmetry has also been suggested in Bose-Einstein condensates \cite{Kle}, a double-well containing the latter with the gain or loss of particles being accounted for simultaneously in one or the other well.

Briefly, in $\mathcal{PT}$-symmetric quantum mechanics, the usual Hermiticity condition is replaced by the commutativity of the Hamiltonian with the product of the parity ($\mathcal{P}: x\mapsto -x, p\mapsto -p, i\mapsto -i$) and time-reversal ($\mathcal{T}: x\mapsto x, p\mapsto -p, i\mapsto -i$) operators

\begin{equation}\label{c1}
\mathcal{PT}H= H\mathcal{PT}
\end{equation}
Such $\mathcal{PT}$-symmetric Hamiltonians may possess either real or conjugate-complex spectra depending on whether all the eigenstates of $H$ are eigenstates of the $\mathcal{PT}$-operator as well (the case of unbroken $\mathcal{PT}$-symmetry) or ceases to be so (the case of broken $\mathcal{PT}$-symmetry).
In general, non-Hermitian Hamiltonians, including a subclass of those coming under the pseudo-Hermitian framework that is supposed to hold the roots of $\mathcal{PT}$, appear in diverse areas of physics such as,  quantum optics, cosmology, atomic and condensed matter physics, magnetohydrodynamics, among others. For detailed reviews of theoretical results and their applications see \cite{mosr,benr} and references therein.

It must however be noted that in contrast to the standard works on $\mathcal{PT}$-symmetry that are largely focussed to linear Schr\"odinger equation, the study of Bose-Einstein condensate. in the mean-field approximation,  comes under the purview of the nonlinear Gross-Pitaevskii equation equipped with the complex potential containing the harmonic-trap condensate.  In nonlinear situations the commutative condition (1.1) is needed to be suitably modified to deal with the Gross-Pitaevskii equation-like equation \cite{wunner}. The effect of the nonlinearity leads to the dramatic feature of the co-existence of $\mathcal{PT}$-symmetric and $\mathcal{PT}$-broken states in certain coupling regions.

 In a recent study of the eigenvalue structure of Bose-Einstein condenstates in a $\mathcal{PT}$-symmetric double well, Dast et al \cite{wunner} identified the category of nonlinear sectors and the idea of $\mathcal{PT}$-symmetry was extended to such systems. As a result of the nonanalytic nonlinear character of the Gross-Pitaevskii equation, the number of solutions does not reflect conservation when the eigenvalue spectrum is confronted with the states undergoing bifurcations even when complex solutions are considered \cite{dast, cart, wunner2, wunner1}. This necessitates an analytic continuation to achieve conservation and the procedure adopted was to separate the Gross-Pitaevskii equation into their real and components. The point is that if one allows complexification for the real and imaginary part of the wave function and the chemical potential, an analytic character of the two coupled equations emerges leading to the conservation of the solutions. Note that in order not to confuse with the usual imaginary unit $i$, another unit of a similar type, the $\hat{i}$, with a property that its square also equals minus one has to be introduced. This results in a transition from the set of complex numbers to an enlarged set of four dimensional hypercomplex numbers, i.e., the bicomplex numbers which are numbers defined by two different complex units. Such a continuation along with the extension of the concept of $\mathcal{PT}$-symmetry onto the domain of bicomplex algebra brings about new symmetries leading to more enriched scenarios and also uncover the existence of new exotic properties in the system. In a way inspired by the work of Dast et al \cite{wunner} we aim to acquire, in the present study, deeper insights into the nature of energy eigenvalues and eigenfunctions of the analogous Schr\"odinger equation (ASI) corresponding to a bicomplex Hamiltonian and investigate the role of the extended $\mathcal{PT}$-symmetry principles. While the work  \cite{wunner} considered the typical two types of  $\mathcal{PT}$-symmetries that bicomplexification offers due to the role of two independent imaginary units namely, $i$  and $\hat{i}$, we explore in this paper the existence of a third type of $\mathcal{PT}$-symmetry resulting from a separate class of the time reversal operator, the $i\hat{i}$, that could flip both the imaginary units. Although handling three different types of distinct $\mathcal{PT}$-symmetries requires lengthy mathematical calculations, nevertheless, their formulation in a single framework provides an exhaustive enquiry into the different roles of $\mathcal{PT}$-symmetry in such a bicomplex manifold. In this connection we must mention that we do not intend to directly study the cases of [26] and [30] but rather want to have an easy and controllable access to a bicomplex system.

In a general way the standard Schr\"odinger Hamiltonian $H(x,p)$ can be bicomplexified considering each physical variable $x$ and $p$ as a bicomplex entity and generalizing the concept of the extended complex phase space by following \cite{kaushal,kp} and defining for instance 
\begin{equation}\label{c1}
\textbf{x}=x_1+ip_2,\qquad \textbf{p}=p_1+ix_2
\end{equation}
where $(x_1,p_1),(x_2,p_2)$ are canonical pairs of phase space variables and $\textbf{p}=-i\hbar\frac{d}{d\textbf{x}}$ is the momentum operator in the usual coordinate space representation satisfying the usual quantum condition
\begin{equation}\label{c2}
\left[\textbf{x},\textbf{p}\right]=i\hbar.
\end{equation}
The Hamiltonian governing such a quantum system can be decomposed as $H(\textbf{x},\textbf{p})=H(x_1,p_1,x_2,p_2)=H_1(x_1,p_1,x_2,p_2)+iH_2(x_1,p_1,x_2,p_2)$.

The plan of this article is as follows: in section \ref{pre} we review the properties of bicomplex numbers, in section \ref{1} we formulate the bicomplex version of the ASE, in section \ref{ext} and section 5 we study the role of different extensions of $\mathcal{PT}$-symmetry defined over the bicomplex algebra carrying out the reduction of the ASE into a system of four coupled partial differential equations whose analytic properties are responsible for the resulting structure of energy eigenvalues and ground-state eigenfunctions, in section \ref{exm} we apply our approach to the typical problems of harmonic oscillator, inverted oscillator and isotonic oscillator ending up finally in section \ref{sum} to present the summary of our findings.
\section{\label{pre}Preliminaries}
\numberwithin{equation}{section}
We note that the set of complex numbers $\mathbb{C}$ consists of elements obtained by duplication of the elements of the set of real numbers $\mathbb{R}$ as induced by a non-real unit $i$ obeying $i^2 =-1$ in the form
\begin{equation}\label{f-dc}
\mathbb{C}=\{z=x+iy:x,y\in\mathbb{R}\}.
\end{equation}
We consider repetition of this duplication process on the members of $\mathbb{C}$ in the presence of a new imaginary unit $\hat{i}$ with the properties
\begin{equation}
{\hat{i}}^2 =-1 ;\quad i\hat{i}= \hat{i}i ; \quad a\hat{i}=\hat{i}a,\forall a \in \mathbb{R}\nonumber
\end{equation}
to extend $\mathbb{C}$ onto the set of bicomplex numbers
\begin{equation}\label{f-db}
\textbf{T} = \{\omega=z_1 +\hat{i}z_2:z_1 ,z_2 \in \mathbb{C}\}.
\end{equation}
In (\ref{f-db}) an additional structure of commutative multiplication is imbedded.

Representing $z_1$ and $z_2$ as defined in (\ref{f-dc}) i.e. $z_1 = x_1 +i x_2$ and $z_2 =x_3 +i x_4$, a bicomplex number acquires the form
\begin{equation}
\omega = x_1 + i x_2 +\hat{i} x_3 +i \hat{i} x_4.
\end{equation}
Clerly $\omega$ is a combination of four units: the unity $1$, two imaginary units $i \mbox { and }\hat{i}$ and one non-real hyperbolic entity $i\hat{i} (=\hat{i}i)$ for which
$(i\hat{i})^2 =1$. In particular if $x_2 =x_3 =0$ the bicomplex number goes over to the hyperbolic number. Looking into the algebraic structure of $\textbf{T}$ it thus becomes a commutative ring with unit.

For two arbitrary bicomplex numbers $\omega= z_1 +\hat{i}z_2$ and $\omega'=z'_1+\hat{i}z'_2$, where $z_1,z_2,z'_1,z'_2\in \mathbb{C}$, the scalar addition and scalar multiplication obey the rules
\begin{eqnarray}
\omega+\omega'&=&(z_1 +z'_1)+\hat{i}(z_2 +z'_2)\\
\omega.\omega'&=&(z_1 z'_1 -z_2 z'_2)+\hat{i}(z_2 z'_1 +z_1 z'_2).
\end{eqnarray}
\subsection{Conjugates and moduli}\label{conj-mod}
Because of the operating of three distinct imaginary units it is evident that a bicomplex number should admit the respective conjugates. Indeed, for any $\omega=z_1 +\hat{i}z_2 \in \textbf{T}$, the possible conjugates are defined as follows
\begin{eqnarray}
\omega^{\dag_1}&=&\bar{z}_1+\hat{i}\bar{z}_2\label{f-con-1}\\
\omega^{\dag_2}&=&z_1-\hat{i}z_2\label{f-con-2}\\
\omega^{\dag_3}&=&\bar{z}_1-\hat{i}\bar{z}_2\label{f-con-3}
\end{eqnarray}
where $\bar{z}_k$ is standard complex conjugation of the complex number $z_k$ implying $\omega^{\dag_3}=(\omega^{\dag_1})^{\dag_2}=(\omega^{\dag_2})^{\dag_1}$.
Each type of conjugation satisfies the standard properties of conjugation:
\begin{eqnarray}
(\omega_1+\omega_2)^{\dag_k}&=&\omega_1^{\dag_k}+\omega_2^{\dag_k}\nonumber\\
(\omega_1^{\dag_k})^{\dag_k}&=&\omega_1\nonumber\\
(\omega_1.\omega_2)^{\dag_k}&=&\omega_1^{\dag_k}.\omega_2^{\dag_k}
\end{eqnarray}
for any $\omega_1,\omega_2\in\textbf{T}$ and $k=1,2,3$.

For the three conjugates, a bicomplex number can have the corresponding three moduli
\begin{eqnarray}
\mid\omega\mid_1^2&=&\omega.\omega^{\dag_2}=z_1^2+z_2^2\label{f-mod-1}\\
\mid\omega\mid_2^2&=&\omega.\omega^{\dag_1}=(\mid z_1\mid^2-\mid z_2\mid^2)+2\hat{i}\mbox{ Re }z_1\bar{z}_2\label{f-mod-2}\\
\mid\omega \mid_3^2&=&\omega.\omega^{\dag_3}=(\mid z_1\mid^2+\mid z_2\mid^2)-2i\hat{i}\mbox{ Im }z_1\bar{z}_2\label{f-mod-3}
\end{eqnarray}
Further the usual Euclidean norm $\parallel.\parallel:\textbf{T}\rightarrow \mathbb{R}$ of $\omega$ reads
\begin{equation}\label{f-norm}
\parallel\omega\parallel = \sqrt{\mid z_1\mid^2+\mid z_2\mid^2}.
\end{equation}
The inverse of $\omega$ is given by
\begin{equation}\label{f-inv}
\omega^{-1}=\frac{\omega^{\dag_2}}{\mid\omega\mid_1^2}.
\end{equation}
If $\omega$ is singular then
\begin{equation}\label{f-sing}
 z_1^2 +  z_2^2 = 0
\end{equation}
holds.
\subsection{Idempotent representation}
 For later use we introduce two bicomplex numbers
 \begin{equation}\label{idcmp}
 \textbf{e}_1 =\frac{1+i\hat{i}}{2},\quad \textbf{e}_2 =\frac{1-i\hat{i}}{2}
 \end{equation}
  satisfying the usual properties
 \begin{equation}
 \textbf{e}_1 +\textbf{e}_2 =1,\quad \textbf{e}_1 . \textbf{e}_2 =\textbf{e}_2 . \textbf{e}_1 =0,\quad \textbf{e}_1^2 =\textbf{e}_1 ,\quad \textbf{e}_2^2 =\textbf{e}_2.
 \end{equation}
      Evidently $\textbf{e}_1 ,\mbox{ and }\textbf{e}_2$ are idempotent. They offer us a unique decomposition of $\textbf{T}$ in that
 for any $\omega = z_1 +\hat{i}z_2 \in \textbf{T}$
  \begin{equation}\label{decomp}
  z_1 +\hat{i}z_2 = (z_1 -iz_2 )\textbf{e}_1 + (z_1 +iz_2 )\textbf{e}_2
  \end{equation}
  is the projection. For a discussion of the unique decomposition of $\omega$ in its idempotent representation, see Appendix B .

\section{\label{1}Analogous Schr\"{o}dinger Equation}
\numberwithin{equation}{section}
To construct a stationary ASE corresponding to the bicomplex Hamiltonian we need to analytically continue (\ref{c1}) on the bicomplex ring $\textbf{T}$ in the manner
\begin{eqnarray}
&& z_1=x_1+ip_1,\qquad p_{z_1}=p_3+ix_3,\nonumber\\
&& z_2=x_4+ip_4,\qquad p_{z_2}=p_2+ix_2\label{bc1}
\end{eqnarray}
where $(x_1,p_3),(x_4,p_2),(x_3,p_1),(x_2,p_4)$ stand for phase space variables.

To deal with the generalization addressing bicomplex numbers, let us consider specifically the decompositions of $\textbf{x}$ and $\textbf{p}$ in the form
\begin{equation}\label{bc-x}
\textbf{x}=z_1+\hat{i}p_{z_2}=x_1+ip_1+\hat{i}p_2+i\hat{i}x_2
\end{equation}
\begin{equation}\label{bc-p}
\textbf{p}=p_{z_1}+\hat{i}z_2=p_3+ix_3+\hat{i}x_4+i\hat{i}p_4,
\end{equation}
where $(x_1,x_2,x_3,x_4)$ are the components in the coordinate space and $(p_1,p_2,p_3,p_4)$ are the components in the momentum space.
These satisfy the generalized commutation relation
\begin{equation}\label{n-comm-rel}
\left[\textbf{x},\textbf{p}\right]=i\hbar \xi I
\end{equation}
$\xi$ being a bicomplex number defined in terms of $\textbf{e}_1$ and $\textbf{e}_2$ namely
\begin{equation}\label{xi}
\xi=\xi_1\textbf{e}_1+\xi_2\textbf{e}_2, 
\end{equation}
where  $\xi_1$ and $\xi_2$ are restricted to be positive quantities \cite{lav}. Note that for the particular case when $\xi_1$ and $\xi_2$ are each  equal to unity, the usual quantum condition is recovered.

From (\ref{n-comm-rel}) using the following representation of the momentum operator 
\begin{equation}\label{bc-comm-rel}
\textbf{p}=-i\hbar\xi\frac{d}{d\textbf{x}}
\end{equation}
we obtain
\begin{equation*}
\frac{d}{d\textbf{x}}=\frac{1}{4}\left[\frac{\partial}{\partial x_1}-i\frac{\partial}{\partial p_1}-\hat{i}\frac{\partial}{\partial p_2}+i\hat{i}\frac{\partial}{\partial x_2}\right]
\end{equation*}
$\textbf{p}$ can be cast in the explicit form
\begin{eqnarray}
\textbf{p}&=& \frac{\hbar}{8}\left\{-(\xi_1+\xi_2)\frac{\partial}{\partial p_1}+(\xi_1-\xi_2)\frac{\partial}{\partial p_2}\right\}
 -i\frac{\hbar}{8}\left\{(\xi_1+\xi_2)\frac{\partial}{\partial x_1}+(\xi_1-\xi_2)\frac{\partial}{\partial x_2}\right\}\nonumber\\
&& +\hat{i}\frac{\hbar}{8}\left\{(\xi_1-\xi_2)\frac{\partial}{\partial x_1}+(\xi_1+\xi_2)\frac{\partial}{\partial x_2}\right\}
 -i\hat{i}\frac{\hbar}{8}\left\{(\xi_1-\xi_2)\frac{\partial}{\partial p_1}-(\xi_1+\xi_2)\frac{\partial}{\partial p_2}\right\}\nonumber\\
&&\label{bc-p-1}
\end{eqnarray}

On comparing with (\ref{bc-p}) we find
\begin{eqnarray}
&& p_3=\frac{\hbar}{8}\left\{-(\xi_1+\xi_2)\frac{\partial}{\partial p_1}+(\xi_1-\xi_2)\frac{\partial}{\partial p_2}\right\},\nonumber\\
&&x_3=-\frac{\hbar}{8}\left\{(\xi_1+\xi_2)\frac{\partial}{\partial x_1}+(\xi_1-\xi_2)\frac{\partial}{\partial x_2}\right\},\nonumber\\
&& x_4=\frac{\hbar}{8}\left\{(\xi_1-\xi_2)\frac{\partial}{\partial x_1}+(\xi_1+\xi_2)\frac{\partial}{\partial x_2}\right\},\nonumber\\
&&p_4=-\frac{\hbar}{8}\left\{(\xi_1-\xi_2)\frac{\partial}{\partial p_1}-(\xi_1+\xi_2)\frac{\partial}{\partial p_2}\right\}.\nonumber
\end{eqnarray}

In the presence of an acting potential $\tilde{V}(\textbf{x})$, the Hamiltonian $H(\textbf{x},\textbf{p})$ reads (in units of $\hbar=m=1$)
\begin{eqnarray}\label{bc-H}
H(\textbf{x},\textbf{p})&=& H(x_1,p_1,p_2,x_2,p_3,x_3,x_4,p_4)\nonumber\\
&=&\frac{\textbf{p}^2}{2}+\tilde{V}(\textbf{x})=-\frac{1}{2}\xi^2\frac{d^2}{d\textbf{x}^2}+\tilde{V}(\textbf{x})
\end{eqnarray}
where
\begin{equation}\label{xi1}
\xi^2=\frac{\xi_1^2+\xi_2^2}{2}+
+i\hat{i}\frac{\xi_1^2-\xi_2^2}{2}.
\end{equation}

For the energy term $\tilde{E}$ and wavefunction $\psi(\textbf{x})\equiv\psi(x_1,p_1,p_2,x_2)$, the ASE then turns out to be as given by
\begin{eqnarray}
&& H\psi(\textbf{x})=\tilde{E}\psi(\textbf{x})\nonumber\\
&\Rightarrow& -\frac{1}{2}\frac{d^2\psi(\textbf{x})}{d\textbf{x}^2}+\frac{1}{16}V(\textbf{x})\psi(\textbf{x})=\frac{1}{16}E\psi(\textbf{x})\label{bc-se}
\end{eqnarray}
where
\begin{equation}\label{bc-ve}
\tilde{V}(\textbf{x})=\frac{1}{16}\xi^2V(\textbf{x}),\quad \tilde{E}=\frac{1}{16}\xi^2E.
\end{equation}

\section{\label{ext}Extended Time Reversal operators}
\numberwithin{equation}{section}
The three types of conjugates of bicomplex numbers introduced earlier admit the corresponding time reversal operators as will be explained in the following discussions.
\subsection{$\mathcal{T}_i$-symmetry}\label{I}
First of all, we observe from the conjugation relation (\ref{f-con-1}) that a class of time reversal operator $\mathcal{T}_i$ can be defined as identified by
\begin{equation}\label{tr-1}
\mathcal{T}_i: i\mapsto -i,\quad \hat{i}\mapsto \hat{i}.
\end{equation}
With the parity operator $\mathcal{P}$ obeying
\begin{eqnarray}
\mathcal{P}&:& x_1\mapsto -x_1\nonumber\\
&& p_1\mapsto -p_1\nonumber\\
&& p_2\mapsto -p_2\nonumber\\
&& x_2\mapsto -x_2\label{par}
\end{eqnarray}
it easily follows from the $\mathcal{PT}$-symmetric character of $\textbf{p}$ i.e. $\mathcal{PT}:\textbf{p}\mapsto \textbf{p}$ that the following features of $\mathcal{T}_i$ hold:
\begin{eqnarray}
\mathcal{T}_i&:& x_1\mapsto x_1\nonumber\\
&& p_1\mapsto -p_1\nonumber\\
&& p_2\mapsto p_2\nonumber\\
&& x_2\mapsto -x_2\label{tr-1-1}
\end{eqnarray}
The restriction on $\xi$ is to be noted:  $\xi_1=\xi_2$. It is also to be remembered that the symmetry operator $\mathcal{PT_i}$ transforms in the coordinate space as $\textbf{x}\mapsto -\textbf{x}$.

\subsection{$\mathcal{T}_{\hat{i}}$-symmetry}\label{II}
While he transformation of $\mathcal{P}$ operator is similar to (\ref{par}), following the conjugation relation (\ref{f-con-2}) we can define a second type of time reversal operator $\mathcal{T}_{\hat{i}}$ that transforms according to
\begin{equation}\label{tr-2}
\mathcal{T}_{\hat{i}}: i\mapsto i,\quad \hat{i}\mapsto -\hat{i}
\end{equation}
As such $\mathcal{T}_{\hat{i}}$ transforms according to
\begin{eqnarray}
\mathcal{T}_{\hat{i}}&:& x_1\mapsto x_1\nonumber\\
&& p_1\mapsto p_1\nonumber\\
&& p_2\mapsto -p_2\nonumber\\
&& x_2\mapsto -x_2\label{tr-2-1}
\end{eqnarray}
with $\xi_1=-\xi_2$. However, because of the positivity restrictions on $\xi_1$ and $\xi_2$ it is clear that the induced $\mathcal{PT}_{\hat{i}}$ cannot be a valid mode of $\mathcal{PT}$ symmetry.

\subsection{$\mathcal{T}_{i\hat{i}}$-symmetry}\label{III}
Finally, following the conjugation relation (\ref{f-con-3}) we can define a third type time reversal operator $\mathcal{T}_{i\hat{i}}$ that undergoes transformations as
\begin{equation}\label{tr-3}
\mathcal{T}_{i\hat{i}}: i\mapsto -i,\quad \hat{i}\mapsto -\hat{i}.
\end{equation}
and may be looked upon \cite{wunner} as the combined operations of $\mathcal{T}_i$ and $\mathcal{T}_{\hat{i}}$  where $\mathcal{T}_i$ again has the physical
interpretation of time reversal and its action in the coordinate space amounts to the replacement  $i\mapsto -i$. Along with (\ref{tr-3}) the transformation properties of $\mathcal{P}$-operator as given by (\ref{par}) hold. Analogously we can define $\mathcal{T}_{\hat{i}}$ as the complex conjugation $\hat{i}\mapsto -\hat{i}$ which, however, has no immediate physical interpretation.

Following the $\mathcal{PT}$-symmetric character of $\textbf{p}$ we also observe that
\begin{eqnarray}
\mathcal{T}_{i\hat{i}}&:& x_1\mapsto x_1\nonumber\\
&& p_1\mapsto -p_1\nonumber\\
&& p_2\mapsto -p_2\nonumber\\
&& x_2\mapsto x_2\label{tr-2-1}
\end{eqnarray}
for all $\xi_1$ and $\xi_2$.

To sum up, we are left with, in a unique way, the pair $\mathcal{PT}_i$ and $\mathcal{PT}_{i\hat{i}}$ as valid candidates  of  $\mathcal{PT}$ symmetry in an extended bicomplex phase space.  In future discussions we therefore address the transformation properties of the underlying systems due to $\mathcal{PT}_i$ and $\mathcal{PT}_{i\hat{i}}$ symmetries. Note that in such a  case it is implied that $\xi_1=\xi_2$. Before we conclude this section it would be worthwhile to look explicitly into the role and interplay of the two above $\mathcal{PT}$-symmetries to classify the different types of eigenvalues associated with them.

For a $\mathcal{PT}_i$-symmetric Hamiltonian $H$
\begin{equation}\label{s1}
[\mathcal{PT}_i ,H]=0,\quad \mathcal{PT}_i \psi=\psi,\quad H\psi=\tilde{E}\psi
\end{equation}
Here unbrokeness of  $\mathcal{PT}_i$ is due to  every eigenfunction $\psi$ of $\mathcal{PT}_i$ symmetric $H$ being  also an eigenfunction of the $\mathcal{PT}_i$ operator. One clearly sees that in such case using the conjugation property defined in (2.6)
\begin{eqnarray}
&& \tilde{E}\psi=H\psi=H\mathcal{PT}_i\psi=\mathcal{PT}_iH\psi=\mathcal{PT}_i\tilde{E}\psi\nonumber\\
&&\Rightarrow [E_1+iE_2+\hat{i}E_3+i\hat{i}E_4]\psi=\mathcal{PT}_i [E_1-iE_2+\hat{i}E_3-i\hat{i}E_4]\psi \nonumber\\
 &&\Rightarrow E_2=E_4=0.\nonumber
 \end{eqnarray}
 
On the other hand, for a $\mathcal{PT}_{i\hat{i}}$-symmetric Hamiltonian $H$
\begin{equation}\label{s1}
[\mathcal{PT}_{i\hat{i}} ,H]=0,\quad \mathcal{PT}_{i\hat{i}} \psi=\psi,\quad H\psi=\tilde{E}\psi
\end{equation}
Here $\mathcal{PT}_{i\hat{i}}$ is unbroken in the sense that if every eigenfunction $\psi$ of $\mathcal{PT}_{i\hat{i}}$ symmetric $H$ is also an eigenfunction of the $\mathcal{PT}_{i\hat{i}}$ operator. One clearly sees that in such case 
\begin{equation}
\tilde{E}\psi=H\psi=H\mathcal{PT}_{i\hat{i}}\psi=\mathcal{PT}_{i\hat{i}}H\psi=\mathcal{PT}_{i\hat{i}}\tilde{E}\psi
\end{equation}
 whence using the conjugation property defined in (2.8) $$[E_1+iE_2+\hat{i}E_3+i\hat{i}E_4]\psi=\mathcal{PT}_{i\hat{i}} [E_1-iE_2-\hat{i}E_3+i\hat{i}E_4]\psi$$  implying   $E_2=E_3=0$.
 This means $\mathcal{PT}_{i}$ and $\mathcal{PT}_{i\hat{i}}$ have to act in conjunction to allow for the
eigenvalues be real.

\section{\label{2}General results}
\numberwithin{equation}{section}
The general four-component form of $V,E,\psi$ can be expanded to be
\begin{eqnarray}
V&=& V_1+iV_2+\hat{i}V_3+i\hat{i}V_4\label{bc-V}\\
E&=& E_1+iE_2+\hat{i}E_3+i\hat{i}E_4\label{bc-E}\\
\psi&=& \psi_1+i\psi_2+\hat{i}\psi_3+i\hat{i}\psi_4\label{bc-psi}
\end{eqnarray}
where $\psi_j,V_j;j=1,2,3,4$ are functions of $x_1,p_1,p_2,x_2$.

Substituting (\ref{bc-V})-(\ref{bc-psi}) into (\ref{bc-se}) a straightforward algebra leads to the following set of relations
\begin{eqnarray}
&&-\frac{1}{2}\left(\frac{\partial^2}{\partial x_1^2}-\frac{\partial^2}{\partial p_1^2}-\frac{\partial^2}{\partial p_2^2}+\frac{\partial^2}{\partial x_2^2}\right)\psi_1-\left(\frac{\partial^2}{\partial x_1 \partial p_1}-\frac{\partial^2}{\partial x_2 \partial p_2}\right)\psi_2\nonumber\\
&&-\left(\frac{\partial^2}{\partial x_1 \partial p_2}-\frac{\partial^2}{\partial x_2 \partial p_1}\right)\psi_3-\left(\frac{\partial^2}{\partial x_1 \partial x_2}+\frac{\partial^2}{\partial p_1 \partial p_2}\right)\psi_4+V_1\psi_1-V_2\psi_2-V_3\psi_3+V_4\psi_4\nonumber\\
&=&E_1\psi_1-E_2\psi_2-E_3\psi_3+E_4\psi_4,\label{bc-1-1}\\
\end{eqnarray}
\begin{eqnarray}
&&\left(\frac{\partial^2}{\partial x_1 \partial p_1}-\frac{\partial^2}{\partial x_2 \partial p_2}\right)\psi_1-\frac{1}{2}\left(\frac{\partial^2}{\partial x_1^2}-\frac{\partial^2}{\partial p_1^2}-\frac{\partial^2}{\partial p_2^2}+\frac{\partial^2}{\partial x_2^2}\right)\psi_2\nonumber\\
&&+\left(\frac{\partial^2}{\partial x_1 \partial x_2}+\frac{\partial^2}{\partial p_1 \partial p_2}\right)\psi_3-\left(\frac{\partial^2}{\partial x_1 \partial p_2}-\frac{\partial^2}{\partial x_2 \partial p_1}\right)\psi_4+V_1\psi_2+V_2\psi_1-V_3\psi_4-V_4\psi_3\nonumber\\
&=&E_1\psi_2+E_2\psi_1-E_3\psi_4-E_4\psi_3,\label{bc-1-2}\\
\end{eqnarray}
\begin{eqnarray}
&&\left(\frac{\partial^2}{\partial x_1 \partial p_2}-\frac{\partial^2}{\partial x_2 \partial p_1}\right)\psi_1+\left(\frac{\partial^2}{\partial x_1 \partial x_2}+\frac{\partial^2}{\partial p_1 \partial p_2}\right)\psi_2-\frac{1}{2}\left(\frac{\partial^2}{\partial x_1^2}-\frac{\partial^2}{\partial p_1^2}-\frac{\partial^2}{\partial p_2^2}+\frac{\partial^2}{\partial x_2^2}\right)\psi_3\nonumber\\
&&-\left(\frac{\partial^2}{\partial x_1 \partial p_1}-\frac{\partial^2}{\partial x_2 \partial p_2}\right)\psi_4+V_1\psi_3-V_2\psi_4+V_3\psi_1-V_4\psi_2\nonumber\\
&=&E_1\psi_3-E_2\psi_4+E_3\psi_1-E_4\psi_2,\label{bc-1-3}\\
\end{eqnarray}
\begin{eqnarray}
&&-\left(\frac{\partial^2}{\partial x_1 \partial x_2}+\frac{\partial^2}{\partial p_1 \partial p_2}\right)\psi_1+\left(\frac{\partial^2}{\partial x_1 \partial p_2}-\frac{\partial^2}{\partial x_2 \partial p_1}\right)\psi_2+\left(\frac{\partial^2}{\partial x_1 \partial p_1}-\frac{\partial^2}{\partial x_2 \partial p_2}\right)\psi_3\nonumber\\
&&-\frac{1}{2}\left(\frac{\partial^2}{\partial x_1^2}-\frac{\partial^2}{\partial p_1^2}-\frac{\partial^2}{\partial p_2^2}+\frac{\partial^2}{\partial x_2^2}\right)\psi_4+V_1\psi_4+V_2\psi_3+V_3\psi_2+V_4\psi_1\nonumber\\
&=&E_1\psi_4+E_2\psi_3+E_3\psi_2+E_4\psi_1.\label{bc-1-4}
\end{eqnarray}

Using the Cauchy-Riemann conditions  \cite{price}:
\begin{eqnarray}
&&\frac{\partial \psi_1}{\partial x_1}=\frac{\partial \psi_2}{\partial p_1}=\frac{\partial \psi_3}{\partial p_2}=\frac{\partial \psi_4}{\partial x_2},\nonumber\\
&&\frac{\partial \psi_2}{\partial x_1}=-\frac{\partial \psi_1}{\partial p_1}=\frac{\partial \psi_4}{\partial p_2}=-\frac{\partial \psi_3}{\partial x_2},\nonumber\\
&&\frac{\partial \psi_3}{\partial x_1}=\frac{\partial \psi_4}{\partial p_1}=-\frac{\partial \psi_1}{\partial p_2}=-\frac{\partial \psi_2}{\partial x_2},\nonumber\\
&&\frac{\partial \psi_4}{\partial x_1}=-\frac{\partial \psi_3}{\partial p_1}=-\frac{\partial \psi_2}{\partial p_2}=\frac{\partial \psi_1}{\partial x_2},\nonumber
\end{eqnarray}
in (\ref{bc-1-1})-(\ref{bc-1-4}) we thus obtain a set of coupled equations
\begin{eqnarray}
&&(\mathfrak{F}+V_1)\psi_1-V_2\psi_2-V_3\psi_3+V_4\psi_4=E_1\psi_1-E_2\psi_2-E_3\psi_3+E_4\psi_4,\nonumber\\
&&(\mathfrak{F}+V_1)\psi_2+V_2\psi_1-V_3\psi_4-V_4\psi_3=E_1\psi_2+E_2\psi_1-E_3\psi_4-E_4\psi_3,\nonumber\\
&&(\mathfrak{F}+V_1)\psi_3-V_2\psi_4+V_3\psi_1-V_4\psi_2=E_1\psi_3-E_2\psi_4+E_3\psi_1-E_4\psi_2,\nonumber\\
&&(\mathfrak{F}+V_1)\psi_4+V_2\psi_3+V_3\psi_2+V_4\psi_1=E_1\psi_4+E_2\psi_3+E_3\psi_2+E_4\psi_1,\label{bc-2-1}
\end{eqnarray}
where
\begin{equation}\label{bc-F}
\mathfrak{F}\equiv -\frac{7}{2}\frac{\partial^2}{\partial x_1^2}+\frac{3}{2}\frac{\partial^2}{\partial p_1^2}+\frac{1}{2}\frac{\partial^2}{\partial p_2^2}-
\frac{5}{2}\frac{\partial^2}{\partial x_2^2}.
\end{equation}
It is convenient to represent (\ref{bc-2-1}) in the matrix form
\begin{equation}\label{bc-2-2}
\mathfrak{M}\Psi=\epsilon \Psi
\end{equation}
where
\begin{equation}\label{bc-2-2-1}
\mathfrak{M}=\left(
  \begin{array}{cccc}
    \mathfrak{F}+V_1 & -V_2 & -V_3 & V_4 \\
    V_2 & \mathfrak{F}+V_1 & -V_4 & -V_3 \\
    V_3 & -V_4 & \mathfrak{F}+V_1 & -V_2 \\
    V_4 & V_3 & V_2 & \mathfrak{F}+V_1 \\
  \end{array}
\right),\quad \epsilon=\left(
                                   \begin{array}{cccc}
                                     E_1 & -E_2 & -E_3 & E_4 \\
                                     E_2 & E_1 & -E_4 & -E_3 \\
                                     E_3 & -E_4 & E_1 & -E_2 \\
                                     E_4 & E_3 & E_2 & E_1 \\
                                   \end{array}
                                 \right),\quad \Psi=\left(
                                                                \begin{array}{c}
                                                                  \psi_1 \\
                                                                  \psi_2 \\
                                                                  \psi_3 \\
                                                                  \psi_4 \\
                                                                \end{array}
                                                              \right).
\end{equation}

We now write down $\mathfrak{M}$ and $\epsilon$ with respect to the idempotent basis matrices $\varepsilon_1,\varepsilon_2$ in their unique representation namely,
\begin{eqnarray}
\mathfrak{M}&=& \varepsilon_1 \mathfrak{M}_1 + \varepsilon_2 \mathfrak{M}_2\label{bc-2-matr-M}\\
 \epsilon &=& \varepsilon_1 \epsilon_1 + \varepsilon_2 \epsilon_2\label{bc-2-matr-eps}
 \end{eqnarray}
  where  $\mathfrak{M}_1,\mathfrak{M}_2$ and $\epsilon_1,\epsilon_2$ stand for
 \begin{eqnarray}
 && \mathfrak{M}_1=\left(
                     \begin{array}{cccc}
                       \mathfrak{F}+V_1+V_4 & -(V_2-V_3) & 0 & 0 \\
                       V_2-V_3 & \mathfrak{F}+V_1+V_4 & 0 & 0 \\
                       0 & 0 & \mathfrak{F}+V_1+V_4 & -(V_2-V_3) \\
                       0 & 0 & V_2-V_3 & \mathfrak{F}+V_1+V_4 \\
                     \end{array}
                   \right),\nonumber\\
                   &&\nonumber\\
                   && \mathfrak{M}_2=\left(
                     \begin{array}{cccc}
                       \mathfrak{F}+V_1-V_4 & -(V_2+V_3) & 0 & 0 \\
                       V_2+V_3 & \mathfrak{F}+V_1-V_4 & 0 & 0 \\
                       0 & 0 & \mathfrak{F}+V_1-V_4 & -(V_2+V_3) \\
                       0 & 0 & V_2+V_3 & \mathfrak{F}+V_1-V_4 \\
                     \end{array}
                   \right),\label{m1,2}\\
                   &&\nonumber\\
 && \epsilon_1=\left(
                 \begin{array}{cccc}
                   E_1+E_4 & -(E_2-E_3) & 0 & 0 \\
                   E_2-E_3 & E_1+E_4 & 0 & 0 \\
                   0 & 0 & E_1+E_4 & -(E_2-E_3) \\
                   0 & 0 & E_2-E_3 & E_1+E_4 \\
                 \end{array}
               \right), \nonumber\\
               &&\nonumber\\
               &&\epsilon_2=\left(
                 \begin{array}{cccc}
                   E_1-E_4 & -(E_2+E_3) & 0 & 0 \\
                   E_2+E_3 & E_1-E_4 & 0 & 0 \\
                   0 & 0 & E_1-E_4 & -(E_2+E_3) \\
                   0 & 0 & E_2+E_3 & E_1-E_4 \\
                 \end{array}
               \right).\label{eps1,2}
 \end{eqnarray}

 Expressing $\Psi=\varepsilon_1(\varepsilon_1 \Psi) +\varepsilon_2(\varepsilon_2 \Psi)$ and substituting (\ref{bc-2-matr-M}) and (\ref{bc-2-matr-eps}) in (\ref{bc-2-2}), we obtain from basiswise comparison
\begin{eqnarray}
\mathfrak{M}_1 \varepsilon_1 \Psi &=& \epsilon_1 \varepsilon_1 \Psi, \label{bc-2-b-1}\\
\mathfrak{M}_2 \varepsilon_2 \Psi &=& \epsilon_2 \varepsilon_2 \Psi. \label{bc-2-b-2}
\end{eqnarray}

To proceed further, we employ (\ref{m1,2}) and (\ref{eps1,2}) to project equation (\ref{bc-2-b-1}) and (\ref{bc-2-b-2}) in their explicit forms
\begin{eqnarray}
&&(\mathfrak{F}+V_1+V_4)(\psi_1+\psi_4)-(V_2-V_3)(\psi_2-\psi_3)=(E_1+E_4)(\psi_1+\psi_4)-(E_2-E_3)(\psi_2-\psi_3),\nonumber\\
&&\label{bc-3-1}\\
&&(\mathfrak{F}+V_1+V_4)(\psi_2-\psi_3)+(V_2-V_3)(\psi_1+\psi_4)=(E_1+E_4)(\psi_2-\psi_3)+(E_2-E_3)(\psi_1+\psi_4),\nonumber\\&&\label{bc-3-2}\\
&&(\mathfrak{F}+V_1-V_4)(\psi_1-\psi_4)-(V_2+V_3)(\psi_2+\psi_3)=(E_1-E_4)(\psi_1-\psi_4)-(E_2+E_3)(\psi_2+\psi_3),\nonumber\\&&\label{bc-3-3}\\
&&(\mathfrak{F}+V_1-V_4)(\psi_2+\psi_3)+(V_2+V_3)(\psi_1-\psi_4)=(E_1-E_4)(\psi_2+\psi_3)+(E_2+E_3)(\psi_1-\psi_4).\nonumber\\&&\label{bc-3-4}
\end{eqnarray}

From  equations (\ref{bc-3-1}) and (\ref{bc-3-2}) we then obtain
\begin{eqnarray}
&&E_1+E_4=\frac{(\psi_1+\psi_4)(\mathfrak{F}+V_1+V_4)(\psi_1+\psi_4)+(\psi_2-\psi_3)(\mathfrak{F}+V_1+V_4)(\psi_2-\psi_3)}{(\psi_1+\psi_4)^2+(\psi_2-\psi_3)^2}
,\label{bc-3-1-1}\\
&&E_2-E_3=\frac{(\psi_1+\psi_4)(\mathfrak{F}+V_1+V_4)(\psi_2-\psi_3)-(\psi_2-\psi_3)(\mathfrak{F}+V_1+V_4)(\psi_1+\psi_4)}{(\psi_1+\psi_4)^2+(\psi_2-\psi_3)^2}
+(V_2-V_3),\nonumber\\&&\label{bc-3-1-2}
\end{eqnarray}
On the other hand, equations (\ref{bc-3-3}) and (\ref{bc-3-4}) give us
\begin{eqnarray}
&&E_1-E_4=\frac{(\psi_1-\psi_4)(\mathfrak{F}+V_1-V_4)(\psi_1-\psi_4)+(\psi_2+\psi_3)(\mathfrak{F}+V_1-V_4)(\psi_2+\psi_3)}{(\psi_1-\psi_4)^2+(\psi_2+\psi_3)^2}
,\label{bc-3-1-3}\\
&&E_2+E_3=\frac{(\psi_1-\psi_4)(\mathfrak{F}+V_1-V_4)(\psi_2+\psi_3)-(\psi_2+\psi_3)(\mathfrak{F}+V_1-V_4)(\psi_1-\psi_4)}{(\psi_1-\psi_4)^2+(\psi_2+\psi_3)^2}
+(V_2+V_3).\nonumber\\&&\label{bc-3-1-4}
\end{eqnarray}

Solving now equations (\ref{bc-3-1-1})-(\ref{bc-3-1-4}) we derive the expressions for $E_i,i=1,2,3,4$ as follows
\begin{eqnarray}
E_1 &=& V_1+\frac{1}{2}[\frac{(\psi_1+\psi_4)\mathfrak{F}(\psi_1+\psi_4)+(\psi_2-\psi_3)\mathfrak{F}(\psi_2-\psi_3)}{(\psi_1+\psi_4)^2+(\psi_2-\psi_3)^2}\nonumber\\
&&+\frac{(\psi_1-\psi_4)\mathfrak{F}(\psi_1-\psi_4)+(\psi_2+\psi_3)\mathfrak{F}(\psi_2+\psi_3)}{(\psi_1-\psi_4)^2+(\psi_2+\psi_3)^2}],\label{bc-3-2-1}\\
&& \nonumber\\
E_2 &=& V_2+\frac{1}{2}[\frac{(\psi_1+\psi_4)\mathfrak{F}(\psi_2-\psi_3)-(\psi_2-\psi_3)\mathfrak{F}(\psi_1+\psi_4)}{(\psi_1+\psi_4)^2+(\psi_2-\psi_3)^2}\nonumber\\
&&+\frac{(\psi_1-\psi_4)\mathfrak{F}(\psi_2+\psi_3)-(\psi_2+\psi_3)\mathfrak{F}(\psi_1-\psi_4)}{(\psi_1-\psi_4)^2+(\psi_2+\psi_3)^2}],\label{bc-3-2-2}\\
&& \nonumber\\
E_3 &=& V_3+\frac{1}{2}[\frac{(\psi_1-\psi_4)\mathfrak{F}(\psi_2+\psi_3)-(\psi_2+\psi_3)\mathfrak{F}(\psi_1-\psi_4)}{(\psi_1-\psi_4)^2+(\psi_2+\psi_3)^2}\nonumber\\
&&-\frac{(\psi_1+\psi_4)\mathfrak{F}(\psi_2-\psi_3)-(\psi_2-\psi_3)\mathfrak{F}(\psi_1+\psi_4)}{(\psi_1+\psi_4)^2+(\psi_2-\psi_3)^2}],\label{bc-3-2-3}\\
&& \nonumber\\
E_4 &=& V_4+\frac{1}{2}[\frac{(\psi_1+\psi_4)\mathfrak{F}(\psi_1+\psi_4)+(\psi_2-\psi_3)\mathfrak{F}(\psi_2-\psi_3)}{(\psi_1+\psi_4)^2+(\psi_2-\psi_3)^2}\nonumber\\
&&-\frac{(\psi_1-\psi_4)\mathfrak{F}(\psi_1-\psi_4)+(\psi_2+\psi_3)\mathfrak{F}(\psi_2+\psi_3)}{(\psi_1-\psi_4)^2+(\psi_2+\psi_3)^2}].\label{bc-3-2-4}
\end{eqnarray}

Making an ansatz for ground state wave function $\psi(x)$:
\begin{equation}\label{wv-1}
\psi(\textbf{x})=e^{g(\textbf{x})}=e^{g_1(\textbf{x})\textbf{e}_1+g_2(\textbf{x})\textbf{e}_2}:\quad g_1=g_{1r}+ig_{1i},g_2=g_{2r}+ig_{2i}
\end{equation}
 (\ref{bc-psi}) gives for each component
\begin{eqnarray}
\psi_1 &=& \frac{1}{2}(e^{g_{1r}}\cos g_{1i}+e^{g_{2r}}\cos g_{2i}),\nonumber\\
\psi_2 &=& \frac{1}{2}(e^{g_{1r}}\sin g_{1i}+e^{g_{2r}}\sin g_{2i}),\nonumber\\
\psi_3 &=& \frac{1}{2}(-e^{g_{1r}}\sin g_{1i}+e^{g_{2r}}\sin g_{2i}),\nonumber\\
\psi_4 &=& \frac{1}{2}(e^{g_{1r}}\cos g_{1i}-e^{g_{2r}}\cos g_{2i}).\label{bc-psi-1}
\end{eqnarray}

The above terms when substituted in (\ref{bc-3-2-1})-(\ref{bc-3-2-4}) gives
\begin{eqnarray}
E_1 &=& V_1+\frac{1}{4}[-7\left\{\frac{\partial^2 g_{1r}}{\partial x_1^2}+(\frac{\partial g_{1r}}{\partial x_1})^2-(\frac{\partial g_{1i}}{\partial x_1})^2\right\}+3\left\{\frac{\partial^2 g_{1r}}{\partial p_1^2}+(\frac{\partial g_{1r}}{\partial p_1})^2-(\frac{\partial g_{1i}}{\partial p_1})^2\right\}\nonumber\\
&&+\left\{\frac{\partial^2 g_{1r}}{\partial p_2^2}+(\frac{\partial g_{1r}}{\partial p_2})^2-(\frac{\partial g_{1i}}{\partial p_2})^2\right\}
-5\left\{\frac{\partial^2 g_{1r}}{\partial x_2^2}+(\frac{\partial g_{1r}}{\partial x_2})^2-(\frac{\partial g_{1i}}{\partial x_2})^2\right\}]\nonumber\\
&+& \frac{1}{4}[-7\left\{\frac{\partial^2 g_{2r}}{\partial x_1^2}+(\frac{\partial g_{2r}}{\partial x_1})^2-(\frac{\partial g_{2i}}{\partial x_1})^2\right\}+3\left\{\frac{\partial^2 g_{2r}}{\partial p_1^2}+(\frac{\partial g_{2r}}{\partial p_1})^2-(\frac{\partial g_{2i}}{\partial p_1})^2\right\}\nonumber\\
&&+\left\{\frac{\partial^2 g_{2r}}{\partial p_2^2}+(\frac{\partial g_{2r}}{\partial p_2})^2-(\frac{\partial g_{2i}}{\partial p_2})^2\right\}
-5\left\{\frac{\partial^2 g_{2r}}{\partial x_2^2}+(\frac{\partial g_{2r}}{\partial x_2})^2-(\frac{\partial g_{2i}}{\partial x_2})^2\right\}].\label{bc-3-3-1}
\end{eqnarray}

\begin{eqnarray}
E_2 &=& V_2+\frac{1}{2}[-7\left\{\frac{1}{2}\frac{\partial^2 g_{2i}}{\partial x_1^2}+\frac{\partial g_{2r}}{\partial x_1}\frac{\partial g_{2i}}{\partial x_1}\right\}
+3\left\{\frac{1}{2}\frac{\partial^2 g_{2i}}{\partial p_1^2}+\frac{\partial g_{2r}}{\partial p_1}\frac{\partial g_{2i}}{\partial p_1}\right\}\nonumber\\
&&+\left\{\frac{1}{2}\frac{\partial^2 g_{2i}}{\partial p_2^2}+\frac{\partial g_{2r}}{\partial p_2}\frac{\partial g_{2i}}{\partial p_2}\right\}-5\left\{\frac{1}{2}\frac{\partial^2 g_{2i}}{\partial x_2^2}+\frac{\partial g_{2r}}{\partial x_2}\frac{\partial g_{2i}}{\partial x_2}\right\}]\nonumber\\
&+&\frac{1}{2}[-7\left\{\frac{1}{2}\frac{\partial^2 g_{1i}}{\partial x_1^2}+\frac{\partial g_{1r}}{\partial x_1}\frac{\partial g_{1i}}{\partial x_1}\right\}
+3\left\{\frac{1}{2}\frac{\partial^2 g_{1i}}{\partial p_1^2}+\frac{\partial g_{1r}}{\partial p_1}\frac{\partial g_{1i}}{\partial p_1}\right\}\nonumber\\
&&+\left\{\frac{1}{2}\frac{\partial^2 g_{1i}}{\partial p_2^2}+\frac{\partial g_{1r}}{\partial p_2}\frac{\partial g_{1i}}{\partial p_2}\right\}-5\left\{\frac{1}{2}\frac{\partial^2 g_{1i}}{\partial x_2^2}+\frac{\partial g_{1r}}{\partial x_2}\frac{\partial g_{1i}}{\partial x_2}\right\}].\label{bc-3-3-2}
\end{eqnarray}

\begin{eqnarray}
E_3 &=& V_3+\frac{1}{2}[-7\left\{\frac{1}{2}\frac{\partial^2 g_{2i}}{\partial x_1^2}+\frac{\partial g_{2r}}{\partial x_1}\frac{\partial g_{2i}}{\partial x_1}\right\}
+3\left\{\frac{1}{2}\frac{\partial^2 g_{2i}}{\partial p_1^2}+\frac{\partial g_{2r}}{\partial p_1}\frac{\partial g_{2i}}{\partial p_1}\right\}\nonumber\\
&&+\left\{\frac{1}{2}\frac{\partial^2 g_{2i}}{\partial p_2^2}+\frac{\partial g_{2r}}{\partial p_2}\frac{\partial g_{2i}}{\partial p_2}\right\}-5\left\{\frac{1}{2}\frac{\partial^2 g_{2i}}{\partial x_2^2}+\frac{\partial g_{2r}}{\partial x_2}\frac{\partial g_{2i}}{\partial x_2}\right\}]\nonumber\\
&-&\frac{1}{2}[-7\left\{\frac{1}{2}\frac{\partial^2 g_{1i}}{\partial x_1^2}+\frac{\partial g_{1r}}{\partial x_1}\frac{\partial g_{1i}}{\partial x_1}\right\}
+3\left\{\frac{1}{2}\frac{\partial^2 g_{1i}}{\partial p_1^2}+\frac{\partial g_{1r}}{\partial p_1}\frac{\partial g_{1i}}{\partial p_1}\right\}\nonumber\\
&&+\left\{\frac{1}{2}\frac{\partial^2 g_{1i}}{\partial p_2^2}+\frac{\partial g_{1r}}{\partial p_2}\frac{\partial g_{1i}}{\partial p_2}\right\}-5\left\{\frac{1}{2}\frac{\partial^2 g_{1i}}{\partial x_2^2}+\frac{\partial g_{1r}}{\partial x_2}\frac{\partial g_{1i}}{\partial x_2}\right\}].\label{bc-3-3-3}
\end{eqnarray}

\begin{eqnarray}
E_4 &=& V_4+\frac{1}{4}[-7\left\{\frac{\partial^2 g_{1r}}{\partial x_1^2}+(\frac{\partial g_{1r}}{\partial x_1})^2-(\frac{\partial g_{1i}}{\partial x_1})^2\right\}+3\left\{\frac{\partial^2 g_{1r}}{\partial p_1^2}+(\frac{\partial g_{1r}}{\partial p_1})^2-(\frac{\partial g_{1i}}{\partial p_1})^2\right\}\nonumber\\
&&+\left\{\frac{\partial^2 g_{1r}}{\partial p_2^2}+(\frac{\partial g_{1r}}{\partial p_2})^2-(\frac{\partial g_{1i}}{\partial p_2})^2\right\}
-5\left\{\frac{\partial^2 g_{1r}}{\partial x_2^2}+(\frac{\partial g_{1r}}{\partial x_2})^2-(\frac{\partial g_{1i}}{\partial x_2})^2\right\}]\nonumber\\
&-& \frac{1}{4}[-7\left\{\frac{\partial^2 g_{2r}}{\partial x_1^2}+(\frac{\partial g_{2r}}{\partial x_1})^2-(\frac{\partial g_{2i}}{\partial x_1})^2\right\}+3\left\{\frac{\partial^2 g_{2r}}{\partial p_1^2}+(\frac{\partial g_{2r}}{\partial p_1})^2-(\frac{\partial g_{2i}}{\partial p_1})^2\right\}\nonumber\\
&&+\left\{\frac{\partial^2 g_{2r}}{\partial p_2^2}+(\frac{\partial g_{2r}}{\partial p_2})^2-(\frac{\partial g_{2i}}{\partial p_2})^2\right\}
-5\left\{\frac{\partial^2 g_{2r}}{\partial x_2^2}+(\frac{\partial g_{2r}}{\partial x_2})^2-(\frac{\partial g_{2i}}{\partial x_2})^2\right\}].\label{bc-3-3-4}
\end{eqnarray}
The above expressions of $E_1,E_2,E_3,E_4$ are the central results of our paper. We now turn to their applications.
\section{\label{exm}Applications}
\numberwithin{equation}{section}
\subsection{\label{HO} Harmonic oscillator}
We first consider the simplest example of the harmonic oscillator whose potential is given by  $\tilde{V}(x)=\frac{1}{16}\xi^2 V(x)$ where
\begin{equation}\label{ho-1}
V(\textbf{x})=a\textbf{x}^2,\quad a>0.
\end{equation}
Then from (\ref{bc-V}), the $V_i$'s $(i=1,2,3,4)$ emerge as
\begin{eqnarray}
V_1&=&a\left(x_1^2-p_1^2-p_2^2+x_2^2\right),\quad V_2=2a\left(x_1p_1-x_2p_2\right),\nonumber\\
V_3&=&2a\left(x_1p_2-x_2p_1\right),\qquad \quad V_4=2a\left(x_1x_2+p_1p_2\right).\nonumber
\end{eqnarray}

In order to determine the energy and ground state wave functions, we express the real and imaginary parts of the  $g_i$'s $(i=1,2)$ appearing in (\ref{wv-1}) as combinations of new quantities $G_i$'s $(i=1,2,3,4)$
\begin{equation}\label{G}
g_{1r}=G_1+G_4,\quad g_{2r}=G_1-G_4,\qquad g_{1i}=G_2-G_3,g_{2i}=G_2+G_3
\end{equation}
where the $G_i$'s are functions of $(x_1,p_1,p_2,x_2)$.

Assuming $G_1$ to be of the form
\begin{equation}\label{G1}
G_1=\alpha (x_1^2-p_1^2-p_2^2+x_2^2)+\beta (x_1p_1-x_2p_2)+\gamma (x_1p_2-x_2p_1)+\delta (x_1x_2+p_1p_2)
\end{equation}
where $\alpha,\beta,\gamma,\delta$ are real constants, and utilizing the Cauchy-Riemann conditions
$$\frac{\partial G_1}{\partial x_1}=\frac{\partial G_2}{\partial p_1};\frac{\partial G_1}{\partial p_1}=-\frac{\partial G_2}{\partial x_1};
\frac{\partial G_1}{\partial p_2}=\frac{\partial G_2}{\partial x_2};\frac{\partial G_1}{\partial x_2}=-\frac{\partial G_2}{\partial p_2}$$ the function $G_2$ gets
restricted to an expression of the type
\begin{equation}\label{G2}
G_2=-\frac{\beta}{2}(x_1^2-p_1^2-p_2^2+x_2^2)+2\alpha (x_1p_1-x_2p_2)-\delta (x_1p_2-x_2p_1)+\gamma (x_1x_2+p_1p_2).
\end{equation}
Similarly the Cauchy-Riemann relations
$$\frac{\partial G_1}{\partial x_1}=\frac{\partial G_3}{\partial p_2};\frac{\partial G_1}{\partial p_1}=\frac{\partial G_3}{\partial x_2};
\frac{\partial G_1}{\partial p_2}=-\frac{\partial G_3}{\partial x_1};\frac{\partial G_1}{\partial x_2}=-\frac{\partial G_3}{\partial p_1}$$
requires $G_3$ to be in the form
\begin{equation}\label{G3}
G_3=-\frac{\gamma}{2}(x_1^2-p_1^2-p_2^2+x_2^2)-\delta (x_1p_1-x_2p_2)+2\alpha (x_1p_2-x_2p_1)+\beta (x_1x_2+p_1p_2).
\end{equation}
Lastly from Cauchy-Riemann relations
$$\frac{\partial G_1}{\partial x_1}=\frac{\partial G_4}{\partial x_2};\frac{\partial G_1}{\partial p_1}=-\frac{\partial G_4}{\partial p_2};
\frac{\partial G_1}{\partial p_2}=-\frac{\partial G_4}{\partial p_1};\frac{\partial G_1}{\partial x_2}=\frac{\partial G_4}{\partial x_1}$$
$G_4$ turns out to be
\begin{equation}\label{G4}
G_4=\frac{\delta}{2}(x_1^2-p_1^2-p_2^2+x_2^2)-\gamma (x_1p_1-x_2p_2)-\beta (x_1p_2-x_2p_1)+2\alpha (x_1x_2+p_1p_2).
\end{equation}

Referring to (\ref{G}), we solve for $g_i$'s to obtain
\begin{eqnarray}
g_{1r}&=&(\alpha+\frac{\delta}{2})(x_1^2-p_1^2-p_2^2+x_2^2)+(\beta-\gamma)(x_1p_1-x_2p_2)-(\beta-\gamma)(x_1p_2-x_2p_1)\nonumber\\
&&+(2\alpha+\delta)(x_1x_2+p_1p_2),\label{g1r}\\
g_{2r}&=&(\alpha-\frac{\delta}{2})(x_1^2-p_1^2-p_2^2+x_2^2)+(\beta+\gamma)(x_1p_1-x_2p_2)+(\beta+\gamma)(x_1p_2-x_2p_1)\nonumber\\
&&-(2\alpha-\delta)(x_1x_2+p_1p_2),\label{g2r}\\
g_{1i}&=&-(\frac{\beta-\gamma}{2})(x_1^2-p_1^2-p_2^2+x_2^2)+(2\alpha+\delta)(x_1p_1-x_2p_2)-(2\alpha+\delta)(x_1p_2-x_2p_1)\nonumber\\
&&-(\beta-\gamma)(x_1x_2+p_1p_2),\label{g1i}\\
g_{2i}&=&-(\frac{\beta+\gamma}{2})(x_1^2-p_1^2-p_2^2+x_2^2)+(2\alpha-\delta)(x_1p_1-x_2p_2)+(2\alpha-\delta)(x_1p_2-x_2p_1)\nonumber\\
&&+(\beta+\gamma)(x_1x_2+p_1p_2).\label{g2i}
\end{eqnarray}

The above solutions enable us to get for the energy values
\begin{eqnarray}
&&E_1=-16\alpha,\quad E_2=8\beta,\quad E_3=8\gamma,\quad E_4=-8\delta ,\label{E-1}
\end{eqnarray}
which are subject to the constraints
\begin{eqnarray}
&&4\alpha^2-\beta^2-\gamma^2+\delta^2=\frac{a}{8},\quad 2\alpha\delta+\beta\gamma=0,\quad
2\alpha\beta-\gamma\delta=0,\quad 2\alpha\gamma-\beta\delta=0.\label{c-1}
\end{eqnarray}
The values of the parameters can be distinguished by two types of results
\begin{eqnarray}
\mbox{Type I}:&& \alpha=\pm \frac{1}{4}\sqrt{\frac{a}{2}},\quad \beta=0,\quad \gamma=0,\quad \delta=0,\nonumber\\
\mbox{Type II}:&& \alpha=0,\quad \beta=0,\quad \gamma=0,\quad \delta=\pm \frac{1}{2}\sqrt{\frac{a}{2}}\nonumber
\end{eqnarray}
signalling the existence of two types of energy values and wave functions. These are summarized below:\\
(a) Type I:
\begin{eqnarray}
E_1&=&\mp 4\sqrt{\frac{a}{2}},\quad E_2=E_3=E_4=0.\label{E-E1-T1}\\
&&\nonumber\\
\psi_1&=&\frac{1}{2}e^{\pm\frac{1}{4}\sqrt{\frac{a}{2}}(x_1^2-p_1^2-p_2^2+x_2^2)}[e^{\pm\frac{1}{2}\sqrt{\frac{a}{2}}(x_1x_2+p_1p_2)}\cos\left\{\pm\frac{1}{2}
\sqrt{\frac{a}{2}}(x_1p_1-x_1p_2+p_1x_2-p_2x_2)\right\}\nonumber\\
&&+e^{\mp\frac{1}{2}\sqrt{\frac{a}{2}}(x_1x_2+p_1p_2)}\cos\left\{\pm\frac{1}{2}
\sqrt{\frac{a}{2}}(x_1p_1+x_1p_2-p_1x_2-p_2x_2)\right\}],\label{psi1-E1-T1}\\
&&\nonumber\\
\psi_2&=&\frac{1}{2}e^{\pm\frac{1}{4}\sqrt{\frac{a}{2}}(x_1^2-p_1^2-p_2^2+x_2^2)}[e^{\pm\frac{1}{2}\sqrt{\frac{a}{2}}(x_1x_2+p_1p_2)}\sin\left\{\pm\frac{1}{2}
\sqrt{\frac{a}{2}}(x_1p_1-x_1p_2+p_1x_2-p_2x_2)\right\}\nonumber\\
&&+e^{\mp\frac{1}{2}\sqrt{\frac{a}{2}}(x_1x_2+p_1p_2)}\sin\left\{\pm\frac{1}{2}
\sqrt{\frac{a}{2}}(x_1p_1+x_1p_2-p_1x_2-p_2x_2)\right\}],\label{psi2-E1-T1}\\
&&\nonumber\\
\psi_3&=&\frac{1}{2}e^{\pm\frac{1}{4}\sqrt{\frac{a}{2}}(x_1^2-p_1^2-p_2^2+x_2^2)}[-e^{\pm\frac{1}{2}\sqrt{\frac{a}{2}}(x_1x_2+p_1p_2)}\sin\left\{\pm\frac{1}{2}
\sqrt{\frac{a}{2}}(x_1p_1-x_1p_2+p_1x_2-p_2x_2)\right\}\nonumber\\
&&+e^{\mp\frac{1}{2}\sqrt{\frac{a}{2}}(x_1x_2+p_1p_2)}\sin\left\{\pm\frac{1}{2}
\sqrt{\frac{a}{2}}(x_1p_1+x_1p_2-p_1x_2-p_2x_2)\right\}],\label{psi3-E1-T1}\\
&&\nonumber\\
\psi_4&=&\frac{1}{2}e^{\pm\frac{1}{4}\sqrt{\frac{a}{2}}(x_1^2-p_1^2-p_2^2+x_2^2)}[e^{\pm\frac{1}{2}\sqrt{\frac{a}{2}}(x_1x_2+p_1p_2)}\cos\left\{\pm\frac{1}{2}
\sqrt{\frac{a}{2}}(x_1p_1-x_1p_2+p_1x_2-p_2x_2)\right\}\nonumber\\
&&-e^{\mp\frac{1}{2}\sqrt{\frac{a}{2}}(x_1x_2+p_1p_2)}\cos\left\{\pm\frac{1}{2}
\sqrt{\frac{a}{2}}(x_1p_1+x_1p_2-p_1x_2-p_2x_2)\right\}].\label{psi4-E1-T1}
\end{eqnarray}
 and
 (b) Type II:
\begin{eqnarray}
&&E_1=E_2=E_3=0,\quad E_4=\mp 4\sqrt{\frac{a}{2}}.\label{E-E1-T2}\\
&&\nonumber\\
\psi_1&=&\frac{1}{2}e^{\pm\frac{1}{2}\sqrt{\frac{a}{2}}(x_1x_2+p_1p_2)}[e^{\pm\frac{1}{4}\sqrt{\frac{a}{2}}(x_1^2-p_1^2-p_2^2+x_2^2)}\cos\left\{\pm\frac{1}{2}
\sqrt{\frac{a}{2}}(x_1p_1-x_1p_2+p_1x_2-p_2x_2)\right\}\nonumber\\
&&+e^{\mp\frac{1}{4}\sqrt{\frac{a}{2}}(x_1^2-p_1^2-p_2^2+x_2^2)}\cos\left\{\mp\frac{1}{2}
\sqrt{\frac{a}{2}}(x_1p_1+x_1p_2-p_1x_2-p_2x_2)\right\}],\label{psi1-E1-T2}\\
&&\nonumber\\
\psi_2&=&\frac{1}{2}e^{\pm\frac{1}{2}\sqrt{\frac{a}{2}}(x_1x_2+p_1p_2)}[e^{\pm\frac{1}{4}\sqrt{\frac{a}{2}}(x_1^2-p_1^2-p_2^2+x_2^2)}\sin\left\{\pm\frac{1}{2}
\sqrt{\frac{a}{2}}(x_1p_1-x_1p_2+p_1x_2-p_2x_2)\right\}\nonumber\\
&&+e^{\mp\frac{1}{4}\sqrt{\frac{a}{2}}(x_1^2-p_1^2-p_2^2+x_2^2)}\sin\left\{\mp\frac{1}{2}
\sqrt{\frac{a}{2}}(x_1p_1+x_1p_2-p_1x_2-p_2x_2)\right\}],\label{psi2-E1-T2}\\
&&\nonumber\\
\psi_3&=&\frac{1}{2}e^{\pm\frac{1}{2}\sqrt{\frac{a}{2}}(x_1x_2+p_1p_2)}[-e^{\pm\frac{1}{4}\sqrt{\frac{a}{2}}(x_1^2-p_1^2-p_2^2+x_2^2)}\sin\left\{\pm\frac{1}{2}
\sqrt{\frac{a}{2}}(x_1p_1-x_1p_2+p_1x_2-p_2x_2)\right\}\nonumber\\
&&+e^{\mp\frac{1}{4}\sqrt{\frac{a}{2}}(x_1^2-p_1^2-p_2^2+x_2^2)}\sin\left\{\mp\frac{1}{2}
\sqrt{\frac{a}{2}}(x_1p_1+x_1p_2-p_1x_2-p_2x_2)\right\}],\label{psi3-E1-T2}\\
&&\nonumber\\
\psi_4&=&\frac{1}{2}e^{\pm\frac{1}{2}\sqrt{\frac{a}{2}}(x_1x_2+p_1p_2)}[e^{\pm\frac{1}{4}\sqrt{\frac{a}{2}}(x_1^2-p_1^2-p_2^2+x_2^2)}\cos\left\{\pm\frac{1}{2}
\sqrt{\frac{a}{2}}(x_1p_1-x_1p_2+p_1x_2-p_2x_2)\right\}\nonumber\\
&&-e^{\mp\frac{1}{4}\sqrt{\frac{a}{2}}(x_1^2-p_1^2-p_2^2+x_2^2)}\cos\left\{\mp\frac{1}{2}
\sqrt{\frac{a}{2}}(x_1p_1+x_1p_2-p_1x_2-p_2x_2)\right\}].\label{psi4-E1-T2}
\end{eqnarray}

Several remarks are in order:\\

Substituting (\ref{E-E1-T1}) into the last relation of (\ref{bc-ve}) we encounter a real energy spectrum for Type I wave function
\begin{equation}\label{en-1-I}
\tilde{E}=\mp\frac{1}{4}\sqrt{\frac{a}{2}}\xi^2.
\end{equation}

However, from (\ref{E-E1-T2}) we are led to a hyperbolic type of energy values for the Type II wave functions
\begin{equation}\label{en-1-II}
\tilde{E}=\mp\frac{1}{4}i\hat{i}\sqrt{\frac{a}{2}}\xi^2.
\end{equation}

Other aspects of our results are as follows:\\

Let us focus on Type I solutions for which there is a real energy spectrum.

$\bullet$ It is easy to see that, under $\mathcal{PT}_i$, $\tilde{V}(x)$ obeys
       $$\mathcal{PT}_i\left(\tilde{V}(\textbf{x})\right)=\mathcal{PT}_i\left(\frac{1}{16}a\xi^2\textbf{x}^2\right)=\frac{1}{16}a\xi^2\textbf{x}^2=\tilde{V}(\textbf{x}),\quad [H,\mathcal{PT}_i]=0.$$

       Following (\ref{psi1-E1-T1})-(\ref{psi4-E1-T1}) since $\mathcal{PT}_i\psi(\textbf{x})=\psi(\textbf{x}),$ it is evident that $\mathcal{PT}_i$-symmetry of $H$ is unbroken.\\

$\bullet$ Further from invariance of $\tilde{V}(x)$ and $\psi(x)$ under $\mathcal{PT}_{i\hat{i}}$ i.e.
       $$\mathcal{PT}_{i\hat{i}}\left(\tilde{V}(\textbf{x})\right)=\tilde{V}(\textbf{x}),\mathcal{PT}_{i\hat{i}}:\psi(\textbf{x})\mapsto \psi(\textbf{x})\quad [H,\mathcal{PT}_{i\hat{i}}]=0$$ it follows that $\mathcal{PT}_{i\hat{i}}$-symmetry of $H$ is unbroken too.\\

A different scenario emerges for Type II solutions:\\

   $\bullet$ While $$\mathcal{PT}_i\left(\tilde{V}(\textbf{x})\right)=\mathcal{PT}_i\left(\frac{1}{16}a\xi^2\textbf{x}^2\right)=\frac{1}{16}a\xi^2\textbf{x}^2=\tilde{V}(\textbf{x}),\quad [H,\mathcal{PT}_i]=0$$ $\psi(\textbf{x})$ does not show the same feature: $$\mathcal{PT}_i\psi(\textbf{x})\neq \lambda \psi(\textbf{x})$$ for any scalar $\lambda$. Hence we conclude that $\mathcal{PT}_i$-symmetry of $H$ is broken.\\

   $\bullet$ Turning to the $\mathcal{PT}_{i\hat{i}}$ operator the situation is slightly different. We have $$\mathcal{PT}_{i\hat{i}}\left(\tilde{V}(\textbf{x})\right)=\tilde{V}(\textbf{x}),\quad [H,\mathcal{PT}_{i\hat{i}}]=0$$ $$\mathcal{PT}_{i\hat{i}}:\psi(\textbf{x})\mapsto \psi(\textbf{x})$$ it follows that $\mathcal{PT}_{i\hat{i}}$-symmetry of $H$ is unbroken.

\subsection{\label{PO} Inverted (Parabolic) oscillator}
For the problem of inverted (parabolic) oscillator acting upon the potential
$\tilde{V}=\frac{1}{16}\xi^2 V$ where
\begin{equation}\label{io-1}
V(\textbf{x})=-b\textbf{x}^2,\quad b>0
\end{equation}
the results for both the classes of solutions reveal the existence of only imaginary energy eigenvalues:
\begin{eqnarray}
&&(a)\quad\mbox{Type I}:\nonumber\\
&& E=\pm4\hat{i}\sqrt{\frac{b}{2}}\quad
\tilde{E}=\mp\frac{1}{4}\hat{i}\sqrt{\frac{b}{2}}\xi^2.\label{E-E2-T1}\\
&&\nonumber\\
\psi_1&=&\frac{1}{2}[e^{\pm\frac{1}{2}\sqrt{\frac{b}{2}}(-x_1p_1+x_1p_2-p_1x_2+p_2x_2)}\cos\left\{\pm \frac{1}{4}\sqrt{\frac{b}{2}}(x_1^2-p_1^2-p_2^2+x_2^2)\pm \frac{1}{2}\sqrt{\frac{b}{2}}(x_1x_2+p_1p_2)\right\}\nonumber\\
&&+e^{\pm\frac{1}{2}\sqrt{\frac{b}{2}}(x_1p_1+x_1p_2-p_1x_2-p_2x_2)}\cos\left\{\mp \frac{1}{4}\sqrt{\frac{b}{2}}(x_1^2-p_1^2-p_2^2+x_2^2)\pm \frac{1}{2}\sqrt{\frac{b}{2}}(x_1x_2+p_1p_2)\right\}],\nonumber\\
&&\label{psi1-E2-T1}\\
\psi_2&=&\frac{1}{2}[e^{\pm\frac{1}{2}\sqrt{\frac{b}{2}}(-x_1p_1+x_1p_2-p_1x_2+p_2x_2)}\sin\left\{\pm \frac{1}{4}\sqrt{\frac{b}{2}}(x_1^2-p_1^2-p_2^2+x_2^2)\pm \frac{1}{2}\sqrt{\frac{b}{2}}(x_1x_2+p_1p_2)\right\}\nonumber\\
&&+e^{\pm\frac{1}{2}\sqrt{\frac{b}{2}}(x_1p_1+x_1p_2-p_1x_2-p_2x_2)}\sin\left\{\mp \frac{1}{4}\sqrt{\frac{b}{2}}(x_1^2-p_1^2-p_2^2+x_2^2)\pm \frac{1}{2}\sqrt{\frac{b}{2}}(x_1x_2+p_1p_2)\right\}],\nonumber\\
&&\label{psi2-E2-T1}\\
\psi_3&=&\frac{1}{2}[-e^{\pm\frac{1}{2}\sqrt{\frac{b}{2}}(-x_1p_1+x_1p_2-p_1x_2+p_2x_2)}\sin\left\{\pm \frac{1}{4}\sqrt{\frac{b}{2}}(x_1^2-p_1^2-p_2^2+x_2^2)\pm \frac{1}{2}\sqrt{\frac{b}{2}}(x_1x_2+p_1p_2)\right\}\nonumber\\
&&+e^{\pm\frac{1}{2}\sqrt{\frac{b}{2}}(x_1p_1+x_1p_2-p_1x_2-p_2x_2)}\sin\left\{\mp \frac{1}{4}\sqrt{\frac{b}{2}}(x_1^2-p_1^2-p_2^2+x_2^2)\pm \frac{1}{2}\sqrt{\frac{b}{2}}(x_1x_2+p_1p_2)\right\}],\nonumber\\
&&\label{psi3-E2-T1}\\
\psi_4&=&\frac{1}{2}[e^{\pm\frac{1}{2}\sqrt{\frac{b}{2}}(-x_1p_1+x_1p_2-p_1x_2+p_2x_2)}\cos\left\{\pm \frac{1}{4}\sqrt{\frac{b}{2}}(x_1^2-p_1^2-p_2^2+x_2^2)\pm \frac{1}{2}\sqrt{\frac{b}{2}}(x_1x_2+p_1p_2)\right\}\nonumber\\
&&-e^{\pm\frac{1}{2}\sqrt{\frac{b}{2}}(x_1p_1+x_1p_2-p_1x_2-p_2x_2)}\cos\left\{\mp \frac{1}{4}\sqrt{\frac{b}{2}}(x_1^2-p_1^2-p_2^2+x_2^2)\pm \frac{1}{2}\sqrt{\frac{b}{2}}(x_1x_2+p_1p_2)\right\}]\nonumber\\
&&\label{psi4-E2-T1}
\end{eqnarray}

\begin{eqnarray}
&&(b)\quad\mbox{Type II}:\nonumber\\
&&E=\pm4i\sqrt{\frac{b}{2}}\quad \tilde{E}=\mp\frac{1}{4}i\sqrt{\frac{b}{2}}\xi^2.\label{E-E2-T2}\\
&&\nonumber\\
\psi_1&=&\frac{1}{2}[e^{\pm\frac{1}{2}\sqrt{\frac{b}{2}}(x_1p_1-x_1p_2+p_1x_2-p_2x_2)}\cos\left\{\mp \frac{1}{4}\sqrt{\frac{b}{2}}(x_1^2-p_1^2-p_2^2+x_2^2)\mp \frac{1}{2}\sqrt{\frac{b}{2}}(x_1x_2+p_1p_2)\right\}\nonumber\\
&&+e^{\pm\frac{1}{2}\sqrt{\frac{b}{2}}(x_1p_1+x_1p_2-p_1x_2-p_2x_2)}\cos\left\{\mp \frac{1}{4}\sqrt{\frac{b}{2}}(x_1^2-p_1^2-p_2^2+x_2^2)\pm \frac{1}{2}\sqrt{\frac{b}{2}}(x_1x_2+p_1p_2)\right\}],\nonumber\\
&&\label{psi1-E2-T2}\\
\psi_2&=&\frac{1}{2}[e^{\pm\frac{1}{2}\sqrt{\frac{b}{2}}(x_1p_1-x_1p_2+p_1x_2-p_2x_2)}\sin\left\{\mp \frac{1}{4}\sqrt{\frac{b}{2}}(x_1^2-p_1^2-p_2^2+x_2^2)\mp \frac{1}{2}\sqrt{\frac{b}{2}}(x_1x_2+p_1p_2)\right\}\nonumber\\
&&+e^{\pm\frac{1}{2}\sqrt{\frac{b}{2}}(x_1p_1+x_1p_2-p_1x_2-p_2x_2)}\sin\left\{\mp \frac{1}{4}\sqrt{\frac{b}{2}}(x_1^2-p_1^2-p_2^2+x_2^2)\pm \frac{1}{2}\sqrt{\frac{b}{2}}(x_1x_2+p_1p_2)\right\}],\nonumber\\
&&\label{psi2-E2-T2}\\
\psi_3&=&\frac{1}{2}[-e^{\pm\frac{1}{2}\sqrt{\frac{b}{2}}(x_1p_1-x_1p_2+p_1x_2-p_2x_2)}\sin\left\{\mp \frac{1}{4}\sqrt{\frac{b}{2}}(x_1^2-p_1^2-p_2^2+x_2^2)\mp \frac{1}{2}\sqrt{\frac{b}{2}}(x_1x_2+p_1p_2)\right\}\nonumber\\
&&+e^{\pm\frac{1}{2}\sqrt{\frac{b}{2}}(x_1p_1+x_1p_2-p_1x_2-p_2x_2)}\sin\left\{\mp \frac{1}{4}\sqrt{\frac{b}{2}}(x_1^2-p_1^2-p_2^2+x_2^2)\pm \frac{1}{2}\sqrt{\frac{b}{2}}(x_1x_2+p_1p_2)\right\}],\nonumber\\
&&\label{psi3-E2-T2}\\
\psi_4&=&\frac{1}{2}[e^{\pm\frac{1}{2}\sqrt{\frac{b}{2}}(x_1p_1-x_1p_2+p_1x_2-p_2x_2)}\cos\left\{\mp \frac{1}{4}\sqrt{\frac{b}{2}}(x_1^2-p_1^2-p_2^2+x_2^2)\mp \frac{1}{2}\sqrt{\frac{b}{2}}(x_1x_2+p_1p_2)\right\}\nonumber\\
&&-e^{\pm\frac{1}{2}\sqrt{\frac{b}{2}}(x_1p_1+x_1p_2-p_1x_2-p_2x_2)}\cos\left\{\mp \frac{1}{4}\sqrt{\frac{b}{2}}(x_1^2-p_1^2-p_2^2+x_2^2)\pm \frac{1}{2}\sqrt{\frac{b}{2}}(x_1x_2+p_1p_2)\right\}].\nonumber\\
&&\label{psi4-E2-T2}
\end{eqnarray}

Corresponding to the above results we find for $\xi_1=\xi_2$ the broken character of $\mathcal{PT}$ to hold\\

                                 $\bullet$ For the $\mathcal{PT}_i$ operator
                                  $$\mathcal{PT}_i\tilde{V}(\textbf{x})=\tilde{V}(\textbf{x}),\quad [H,\mathcal{PT}_i]=0$$ where $\mathcal{PT}_i\psi(\textbf{x})\neq \lambda \psi(\textbf{x})$ for any scalar $\lambda$ and so $\mathcal{PT}_i$-symmetry of $H$ is broken.\\

                                 $\bullet$ For the $\mathcal{PT}_{i\hat{i}}$ operator
                                 $$\mathcal{PT}_{i\hat{i}}\tilde{V}(\textbf{x})=\tilde{V}(\textbf{x}),\quad [H,\mathcal{PT}_{i\hat{i}}]=0$$ where $\mathcal{PT}_{i\hat{i}}\psi(\textbf{x})\neq \lambda \psi(\textbf{x})$ for any scalar $\lambda$ and so $\mathcal{PT}_{i\hat{i}}$-symmetry of $H$ is broken.

\subsection{\label{IO} Isotonic oscillator}
We now address the problem of isotonic oscillator which is governed by the potential $\tilde{V}=\frac{1}{16}\xi^2 V$ where
\begin{equation}\label{iso1}
V(\textbf{x})=a{\textbf{x}}^2+\frac{b}{{\textbf{x}}^2},\quad a(\neq 0),b(\neq 0).
\end{equation}
From (\ref{bc-V}) we then obtain
\begin{eqnarray}
V_1 &=& a\left(x_1^2-p_1^2-p_2^2+x_2^2\right)+\frac{b}{2}\left[\frac{(x_1+x_2)^2-(p_1-p_2)^2}{\{(x_1+x_2)^2+(p_1-p_2)^2\}^2}
+\frac{(x_1-x_2)^2-(p_1+p_2)^2}{\{(x_1-x_2)^2+(p_1+p_2)^2\}^2}\right],\nonumber\\
V_2 &=& 2a\left(x_1p_1-x_2p_2\right)-b\left[\frac{(x_1+x_2)(p_1-p_2)}{\{(x_1+x_2)^2+(p_1-p_2)^2\}^2}
+\frac{(x_1-x_2)(p_1+p_2)}{\{(x_1-x_2)^2+(p_1+p_2)^2\}^2}\right],\nonumber\\
V_3 &=& 2a\left(x_1p_2-x_2p_1\right)+b\left[\frac{(x_1+x_2)(p_1-p_2)}{\{(x_1+x_2)^2+(p_1-p_2)^2\}^2}
-\frac{(x_1-x_2)(p_1+p_2)}{\{(x_1-x_2)^2+(p_1+p_2)^2\}^2}\right],\nonumber\\
V_4 &=& 2a\left(x_1x_2+p_1p_2\right)+\frac{b}{2}\left[\frac{(x_1+x_2)^2-(p_1-p_2)^2}{\{(x_1+x_2)^2+(p_1-p_2)^2\}^2}
-\frac{(x_1-x_2)^2-(p_1+p_2)^2}{\{(x_1-x_2)^2+(p_1+p_2)^2\}^2}\right].\nonumber
\end{eqnarray}

By inspection of the terms $V_1,V_2,V_3,V_4$  that we consider the following ansatz for $G_1$
\begin{eqnarray}
G_1&=& \alpha_1(x_1^2-p_1^2-p_2^2+x_2^2)+\alpha_2(x_1p_1-x_2p_2)+\alpha_3(x_1p_2-x_2p_1)+\alpha_4(x_1x_2+p_1p_2)\nonumber\\
&+&\beta_1 \tan^{-1}(\frac{x_1+x_2}{p_1-p_2})
+\beta_2 \tan^{-1}(\frac{x_1-x_2}{p_1+p_2})+\beta_3 \log\{(x_1+x_2)^2+(p_1-p_2)^2\}\nonumber\\
&+&\beta_4 \log\{(x_1-x_2)^2+(p_1+p_2)^2\}\label{iso-G1}
\end{eqnarray}
for real constants $\alpha_i,\beta_i : i=1,2,3,4$.

Then proceeding in the same way as followed in the case of the harmonic oscillator we find
\begin{eqnarray}
G_2&=& -\frac{\alpha_2}{2}(x_1^2-p_1^2-p_2^2+x_2^2)+2\alpha_1(x_1p_1-x_2p_2)-\alpha_4(x_1p_2-x_2p_1)+\alpha_3(x_1x_2+p_1p_2)\nonumber\\
&-&2\beta_3 \tan^{-1}(\frac{x_1+x_2}{p_1-p_2})
-2\beta_4 \tan^{-1}(\frac{x_1-x_2}{p_1+p_2})+\frac{\beta_1}{2} \log\{(x_1+x_2)^2+(p_1-p_2)^2\}\nonumber\\
&+&\frac{\beta_2}{2} \log\{(x_1-x_2)^2+(p_1+p_2)^2\},\label{iso-G2}\\\nonumber\\
G_3&=& -\frac{\alpha_3}{2}(x_1^2-p_1^2-p_2^2+x_2^2)-\alpha_4(x_1p_1-x_2p_2)+2\alpha_1(x_1p_2-x_2p_1)+\alpha_2(x_1x_2+p_1p_2)\nonumber\\
&+&2\beta_3\tan^{-1}(\frac{x_1+x_2}{p_1-p_2})
-2\beta_4 \tan^{-1}(\frac{x_1-x_2}{p_1+p_2})-\frac{\beta_1}{2}\log\{(x_1+x_2)^2+(p_1-p_2)^2\}\nonumber\\
&+&\frac{\beta_2}{2}\log\{(x_1-x_2)^2+(p_1+p_2)^2\},\label{iso-G3}\\\nonumber\\
G_4&=& \frac{\alpha_4}{2}(x_1^2-p_1^2-p_2^2+x_2^2)-\alpha_3(x_1p_1-x_2p_2)-\alpha_2(x_1p_2-x_2p_1)+2\alpha_1(x_1x_2+p_1p_2)\nonumber\\
&+&\beta_1 \tan^{-1}(\frac{x_1+x_2}{p_1-p_2})
-\beta_2 \tan^{-1}(\frac{x_1-x_2}{p_1+p_2})+\beta_3\log\{(x_1+x_2)^2+(p_1-p_2)^2\}\nonumber\\
&-&\beta_4\log\{(x_1-x_2)^2+(p_1+p_2)^2\}.\label{iso-G4}
\end{eqnarray}
These expressions for $G_i$'s $(i=1,2,3,4)$ yield
\begin{eqnarray}
g_{1r}&=&(\alpha_1+\frac{\alpha_4}{2})(x_1^2-p_1^2-p_2^2+x_2^2)+(\alpha_2-\alpha_3)(x_1p_1-x_2p_2-x_1p_2+x_2p_1)\nonumber\\
&+&(2\alpha_1+\alpha_4)(x_1x_2+p_1p_2)
+2\beta_1 \tan^{-1}(\frac{x_1+x_2}{p_1-p_2})\nonumber\\
&+&2\beta_3 \log\{(x_1+x_2)^2+(p_1-p_2)^2\},\label{iso-g1r}\\\nonumber\\
g_{1i}&=&-\frac{1}{2}(\alpha_2-\alpha_3)(x_1^2-p_1^2-p_2^2+x_2^2)+(2\alpha_1+\alpha_4)(x_1p_1-x_2p_2-x_1p_2+x_2p_1)\nonumber\\
&-&(\alpha_2-\alpha_3)(x_1x_2+p_1p_2)-4\beta_3 \tan^{-1}(\frac{x_1+x_2}{p_1-p_2})\nonumber\\
&+&\beta_1 \log\{(x_1+x_2)^2+(p_1-p_2)^2\},\label{iso-g1i}\\\nonumber\\
g_{2r}&=&(\alpha_1-\frac{\alpha_4}{2})(x_1^2-p_1^2-p_2^2+x_2^2)+(\alpha_2+\alpha_3)(x_1p_1-x_2p_2+x_1p_2-x_2p_1)\nonumber\\
&-&(2\alpha_1-\alpha_4)(x_1x_2+p_1p_2)+2\beta_2 \tan^{-1}(\frac{x_1-x_2}{p_1+p_2})\nonumber\\
&+&2\beta_4 \log\{(x_1-x_2)^2+(p_1+p_2)^2\},\label{iso-g2r}\\\nonumber\\
g_{2i}&=&-\frac{1}{2}(\alpha_2+\alpha_3)(x_1^2-p_1^2-p_2^2+x_2^2)+(2\alpha_1-\alpha_4)(x_1p_1-x_2p_2+x_1p_2-x_2p_1)\nonumber\\
&+&(\alpha_2+\alpha_3)(x_1x_2+p_1p_2)-4\beta_4 \tan^{-1}(\frac{x_1-x_2}{p_1+p_2})\nonumber\\
&+&\beta_2 \log\{(x_1-x_2)^2+(p_1+p_2)^2\}.\label{iso-g2i}
\end{eqnarray}

Substituting ing the above solutions in (\ref{bc-3-3-1})-(\ref{bc-3-3-4}) we get
\begin{eqnarray}
&&\alpha_2=\alpha_3=0 \label{IO-1}\\
&&4\alpha_1^2+\alpha_4^2=\frac{a}{8},\qquad \alpha_1\alpha_4=0 \label{IO-2}\\
&&E_1=-4\left[(1+8\beta_3)(2\alpha_1+\alpha_4)+(1+8\beta_4)(2\alpha_1-\alpha_4)\right]\label{IO-3}\\
&&E_2=-16\beta_1(2\alpha_1+\alpha_4)-16\beta_2(2\alpha_1-\alpha_4)\label{IO-4}\\
&&E_3=16\beta_1(2\alpha_1+\alpha_4)-16\beta_2(2\alpha_1-\alpha_4)\label{IO-5}\\
&&E_4=-4\left[(1+8\beta_3)(2\alpha_1+\alpha_4)-(1+8\beta_4)(2\alpha_1-\alpha_4)\right]\label{IO-6}\\
&&4\beta_3^2-\beta_1^2-\beta_3=4\beta_4^2-\beta_2^2-\beta_4=b\label{IO-7}\\
&&\beta_1(8\beta_3-1)=\beta_2(8\beta_4-1)=0\label{IO-8}
\end{eqnarray}
A plausible set of viable solutions for the real parameters $\alpha_i,\beta_i:i=1,2,3,4$ is given by
\begin{eqnarray}\label{6.3.10}
&&\beta_1=\beta_2=0,\quad \Rightarrow E_2=E_3=0 \label{IO-9}\\
&&\beta_3=\frac{1\pm\sqrt{1+\frac{b}{2}}}{8}\label{IO-10a}\\
&&\beta_4=\frac{1\pm\sqrt{1+\frac{b}{2}}}{8}\label{IO-10b}
\end{eqnarray}
for the following restrictions on the real coupling constants $a$ and $b$ of the isotonic oscillator potential:
\begin{equation}\label{iso-par}
a>0 \mbox{ and } b\geq -2.
\end{equation}

We thus have two types of solutions for the parameters:
\begin{eqnarray}
\mbox{Type I}:&& \alpha_1=\pm \frac{\sqrt{a}}{4\sqrt{2}},\quad \alpha_2=\alpha_3=\alpha_4=\beta_1=\beta_2=0,\nonumber\\
\mbox{Type II}:&& \alpha_4=\pm \frac{\sqrt{a}}{2\sqrt{2}},\quad \alpha_1=\alpha_2=\alpha_3=\beta_1=\beta_2=0.\nonumber
\end{eqnarray}

Consequently two types of energy values and energy eigenfunctions emerge. These are given as follows:\\
\indent
(a)Type I:\quad The results are
\begin{eqnarray}
&&E_1=-16\alpha_1[1+4(\beta_3+\beta_4)],\quad E_4=0 \Rightarrow \tilde{E}=\frac{1}{16}\xi^2E_1\nonumber
\end{eqnarray}
\begin{eqnarray}
\psi_1&=&\frac{1}{2}e^{\alpha_1(x_1^2-p_1^2-p_2^2+x_2^2)+2\alpha_1(x_1x_2+p_1p_2)+2\beta_3\log\{(x_1+x_2)^2+(p_1-p_2)^2\}}\nonumber\\
&&.\cos\{2\alpha_1(x_1p_1-x_2p_2-x_1p_2+x_2p_1)-4\beta_3\tan^{-1}(\frac{x_1+x_2}{p_1-p_2})\}\nonumber\\
&&+\frac{1}{2}e^{\alpha_1(x_1^2-p_1^2-p_2^2+x_2^2)-2\alpha_1(x_1x_2+p_1p_2)+2\beta_4\log\{(x_1-x_2)^2+(p_1+p_2)^2\}}\nonumber\\
&&.\cos\{2\alpha_1(x_1p_1-x_2p_2+x_1p_2-x_2p_1)-4\beta_4\tan^{-1}(\frac{x_1-x_2}{p_1+p_2})\}\nonumber\\
\end{eqnarray}
\begin{eqnarray}
\psi_2&=&\frac{1}{2}e^{\alpha_1(x_1^2-p_1^2-p_2^2+x_2^2)+2\alpha_1(x_1x_2+p_1p_2)+2\beta_3\log\{(x_1+x_2)^2+(p_1-p_2)^2\}}\nonumber\\
&&.\sin\{2\alpha_1(x_1p_1-x_2p_2-x_1p_2+x_2p_1)-4\beta_3\tan^{-1}(\frac{x_1+x_2}{p_1-p_2})\}\nonumber\\
&&+\frac{1}{2}e^{\alpha_1(x_1^2-p_1^2-p_2^2+x_2^2)-2\alpha_1(x_1x_2+p_1p_2)+2\beta_4\log\{(x_1-x_2)^2+(p_1+p_2)^2\}}\nonumber\\
&&.\sin\{2\alpha_1(x_1p_1-x_2p_2+x_1p_2-x_2p_1)-4\beta_4\tan^{-1}(\frac{x_1-x_2}{p_1+p_2})\}\nonumber\\
\end{eqnarray}
\begin{eqnarray}
\psi_3&=&-\frac{1}{2}e^{\alpha_1(x_1^2-p_1^2-p_2^2+x_2^2)+2\alpha_1(x_1x_2+p_1p_2)+2\beta_3\log\{(x_1+x_2)^2+(p_1-p_2)^2\}}\nonumber\\
&&.\sin\{2\alpha_1(x_1p_1-x_2p_2-x_1p_2+x_2p_1)-4\beta_3\tan^{-1}(\frac{x_1+x_2}{p_1-p_2})\}\nonumber\\
&&+\frac{1}{2}e^{\alpha_1(x_1^2-p_1^2-p_2^2+x_2^2)-2\alpha_1(x_1x_2+p_1p_2)+2\beta_4\log\{(x_1-x_2)^2+(p_1+p_2)^2\}}\nonumber\\
&&.\sin\{2\alpha_1(x_1p_1-x_2p_2+x_1p_2-x_2p_1)-4\beta_4\tan^{-1}(\frac{x_1-x_2}{p_1+p_2})\}\nonumber\\
\end{eqnarray}
\begin{eqnarray}
\psi_4&=&\frac{1}{2}e^{\alpha_1(x_1^2-p_1^2-p_2^2+x_2^2)+2\alpha_1(x_1x_2+p_1p_2)+2\beta_3\log\{(x_1+x_2)^2+(p_1-p_2)^2\}}\nonumber\\
&&.\cos\{2\alpha_1(x_1p_1-x_2p_2-x_1p_2+x_2p_1)-4\beta_3\tan^{-1}(\frac{x_1+x_2}{p_1-p_2})\}\nonumber\\
&&-\frac{1}{2}e^{\alpha_1(x_1^2-p_1^2-p_2^2+x_2^2)-2\alpha_1(x_1x_2+p_1p_2)+2\beta_4\log\{(x_1-x_2)^2+(p_1+p_2)^2\}}\nonumber\\
&&.\cos\{2\alpha_1(x_1p_1-x_2p_2+x_1p_2-x_2p_1)-4\beta_4\tan^{-1}(\frac{x_1-x_2}{p_1+p_2})\}\nonumber\\
\end{eqnarray}

For the extended $\mathcal{PT}$-symmetry for Type I solutions along with  $\beta_3=\beta_4$ we therefore find the typical unbroken character of $\mathcal{PT}$ :\\

                                 $\bullet$ $$\mathcal{PT}_i\tilde{V}(\textbf{x})=\tilde{V}(\textbf{x}),\quad [H,\mathcal{PT}_i]=0$$ where $\mathcal{PT}_i\psi(\textbf{x})=\psi(\textbf{x})$ and so $\mathcal{PT}_i$-symmetry of $H$ is unbroken.\\

                                 $\bullet$ $$\mathcal{PT}_{i\hat{i}}\tilde{V}(\textbf{x})=\tilde{V}(\textbf{x}),\quad [H,\mathcal{PT}_{i\hat{i}}]=0$$ where $\mathcal{PT}_{i\hat{i}}\psi(\textbf{x})=\psi(\textbf{x})$ so that $\mathcal{PT}_{i\hat{i}}$-symmetry of $H$ is unbroken.\\

However even if $\beta_3\neq\beta_4$ the unbroken character of $\mathcal{PT}_{i\hat{i}}$ does not change. But the scenario is different for $\mathcal{PT}_i$ and $\mathcal{PT}_{\hat{i}}$ operators as both of them are broken for $\beta_3\neq\beta_4$.\\
\indent
(b) Type II: \quad The results are
\begin{eqnarray}
&&E_1=0,\quad E_4=-8\alpha_4[1+4(\beta_3+\beta_4)] \Rightarrow \tilde{E}=\frac{1}{16}i\hat{i}\xi^2 E_4\nonumber
\end{eqnarray}
\begin{eqnarray}
\psi_1&=&\frac{1}{2}e^{\frac{\alpha_4}{2}(x_1^2-p_1^2-p_2^2+x_2^2)+\alpha_4(x_1x_2+p_1p_2)+2\beta_3\log\{(x_1+x_2)^2+(p_1-p_2)^2\}}\nonumber\\
&&.\cos\{\alpha_4(x_1p_1-x_2p_2-x_1p_2+x_2p_1)-4\beta_3\tan^{-1}(\frac{x_1+x_2}{p_1-p_2})\}\nonumber\\
&&+\frac{1}{2}e^{-\frac{\alpha_4}{2}(x_1^2-p_1^2-p_2^2+x_2^2)+\alpha_4(x_1x_2+p_1p_2)+2\beta_4\log\{(x_1-x_2)^2+(p_1+p_2)^2\}}\nonumber\\
&&.\cos\{-\alpha_4(x_1p_1-x_2p_2+x_1p_2-x_2p_1)-4\beta_4\tan^{-1}(\frac{x_1-x_2}{p_1+p_2})\}\nonumber\\
\end{eqnarray}
\begin{eqnarray}
\psi_2&=&\frac{1}{2}e^{\frac{\alpha_4}{2}(x_1^2-p_1^2-p_2^2+x_2^2)+\alpha_4(x_1x_2+p_1p_2)+2\beta_3\log\{(x_1+x_2)^2+(p_1-p_2)^2\}}\nonumber\\
&&.\sin\{\alpha_4(x_1p_1-x_2p_2-x_1p_2+x_2p_1)-4\beta_3\tan^{-1}(\frac{x_1+x_2}{p_1-p_2})\}\nonumber\\
&&+\frac{1}{2}e^{-\frac{\alpha_4}{2}(x_1^2-p_1^2-p_2^2+x_2^2)+\alpha_4(x_1x_2+p_1p_2)+2\beta_4\log\{(x_1-x_2)^2+(p_1+p_2)^2\}}\nonumber\\
&&.\sin\{-\alpha_4(x_1p_1-x_2p_2+x_1p_2-x_2p_1)-4\beta_4\tan^{-1}(\frac{x_1-x_2}{p_1+p_2})\}\nonumber\\
\end{eqnarray}
\begin{eqnarray}
\psi_3&=&-\frac{1}{2}e^{\frac{\alpha_4}{2}(x_1^2-p_1^2-p_2^2+x_2^2)+\alpha_4(x_1x_2+p_1p_2)+2\beta_3\log\{(x_1+x_2)^2+(p_1-p_2)^2\}}\nonumber\\
&&.\sin\{\alpha_4(x_1p_1-x_2p_2-x_1p_2+x_2p_1)-4\beta_3\tan^{-1}(\frac{x_1+x_2}{p_1-p_2})\}\nonumber\\
&&+\frac{1}{2}e^{-\frac{\alpha_4}{2}(x_1^2-p_1^2-p_2^2+x_2^2)+\alpha_4(x_1x_2+p_1p_2)+2\beta_4\log\{(x_1-x_2)^2+(p_1+p_2)^2\}}\nonumber\\
&&.\sin\{-\alpha_4(x_1p_1-x_2p_2+x_1p_2-x_2p_1)-4\beta_4\tan^{-1}(\frac{x_1-x_2}{p_1+p_2})\}\nonumber\\
\end{eqnarray}
\begin{eqnarray}
\psi_4&=&\frac{1}{2}e^{\frac{\alpha_4}{2}(x_1^2-p_1^2-p_2^2+x_2^2)+\alpha_4(x_1x_2+p_1p_2)+2\beta_3\log\{(x_1+x_2)^2+(p_1-p_2)^2\}}\nonumber\\
&&.\cos\{\alpha_4(x_1p_1-x_2p_2-x_1p_2+x_2p_1)-4\beta_3\tan^{-1}(\frac{x_1+x_2}{p_1-p_2})\}\nonumber\\
&&-\frac{1}{2}e^{-\frac{\alpha_4}{2}(x_1^2-p_1^2-p_2^2+x_2^2)+\alpha_4(x_1x_2+p_1p_2)+2\beta_4\log\{(x_1-x_2)^2+(p_1+p_2)^2\}}\nonumber\\
&&.\cos\{-\alpha_4(x_1p_1-x_2p_2+x_1p_2-x_2p_1)-4\beta_4\tan^{-1}(\frac{x_1-x_2}{p_1+p_2})\}\nonumber\\\nonumber\\\nonumber\\
\end{eqnarray}

Hence for Type II solutions our conclusions are that whatever the values of $\beta_3$ and $\beta_4$, the $\mathcal{PT}$-symmetry works in a different way:\\

                                 $\bullet$ $$\mathcal{PT}_i\tilde{V}(\textbf{x})=\tilde{V}(\textbf{x}),\quad [H,\mathcal{PT}_i]=0$$ where $\mathcal{PT}_i\psi(\textbf{x})\neq \lambda \psi(\textbf{x})$ for any scalar $\lambda$ and so $\mathcal{PT}_i$-symmetry of $H$ is broken.\\

                                $\bullet$  $$\mathcal{PT}_{i\hat{i}}\tilde{V}(\textbf{x})=\tilde{V}(\textbf{x}),\quad [H,\mathcal{PT}_{i\hat{i}}]=0$$ and $\mathcal{PT}_{i\hat{i}}\psi(\textbf{x})=\psi(\textbf{x})$  it follows that $\mathcal{PT}_{i\hat{i}}$-symmetry of $H$ is unbroken.
                                
Finally in this problem of isotonic oscillator, extension of the real coupling constant $b$ to its bicomplex counterpart, reveals an interesting feature. To  fit into our formalism the restrictions $b_2=b_3=0$ and $\mid b_4 \mid \leq (2+b_1), b_1\geq -2$ were required to be imposed upon the coupling constants. If $b_4\neq0,\alpha_4=0$, although the ground state energy $\tilde{E}$ is real and $\mathcal{PT}_i\tilde{V}(\textbf{x})=\tilde{V}(\textbf{x})$ along with $[H,\mathcal{PT}_i]=0$,the $\mathcal{PT}_i$-symmetry is broken since $\mathcal{PT}_i\psi(\textbf{x})\neq \lambda \psi(\textbf{x})$ for any scalar $\lambda$. But the scenario is completely different for the $\mathcal{PT}_{i\hat{i}}$ operator: as  $\mathcal{PT}_{i\hat{i}}\tilde{V}(\textbf{x})=\tilde{V}(\textbf{x})$ and $[H,\mathcal{PT}_{i\hat{i}}]=0$ as well as $\mathcal{PT}_{i\hat{i}}\psi(\textbf{x})=\psi(\textbf{x})$, the $\mathcal{PT}_{i\hat{i}}$-symmetry remains unbroken.

\section{\label{sum}Summary}
To summarize we took up in this paper a quantitative analysis of bicomplex algebra that leads to associated Hamiltonians couched in an analogous version of the Schr\"odinger equation. Bicomplex numbers being basically four dimensional hypercomplex numbers can admit of different types of conjugation each defining a separate class of the time reversal operator. As a result we could set up different extensions of parity ($\mathcal{P}$)-time ($\mathcal{T}$)-symmetric models such as the ones corresponding to $\mathcal{PT}_i$, $\mathcal{PT}_{\hat{i}}$ and $\mathcal{PT}_{i\hat{i}}$ operators in an extended phase space formalism. However, as we have explicitly demonstrated, $\mathcal{PT}_{\hat{i}}$ is not a valid candidate for a $\mathcal{PT}$-symmetric operator. By writing down suitable representations and exploiting the Cauchy-Riemann conditions judiciously we showed that we could arrive at the closed-form expressions of the energy and wave function components for a given choice of the potential.   Our procedure was then applied to the problems of harmonic oscillator, inverted oscillator and isotonic oscillator. In all such systems we obtained two types of solutions each revealing specific $\mathcal{PT}$ properties. In particular we observed that a real energy value exists for all the three cases when the Hamiltonian $H$ obeys unbroken $\mathcal{PT}_i$ and $\mathcal{PT}_{i\hat{i}}$-symmetries.  

\section*{Acknowledgement}
One of us (AB) gratefully acknowledges University Grants Commission, New Delhi for awarding a minor research project. We thank the anonymous referees for their criticisms and making constructive suggestions that have helped in the improvement of the paper.

\newpage
\begin{center}
\textbf{Appendix-A}
\end{center}
\textbf{Bicomplex valued functions:}\\
Any bicomplex function $f:\Omega \subset \textbf{T} \mapsto \textbf{T}$ involving unique idempotent decomposition into two complex valued functions reads
\begin{equation}\label{f-fn}
f(\omega)=f_1 (\omega_1 )\textbf{e}_1 + f_2 (\omega_2 )\textbf{e}_2 ,\quad \omega = (\omega_1 \textbf{e}_1 +\omega_2 \textbf{e}_2)\in\Omega
\end{equation}
where $\omega_1\in\Omega_1,\omega_2\in\Omega_2$ and $\Omega=\Omega_1\textbf{e}_1+\Omega_2\textbf{e}_2$.

The derivative of $f$ at a point $\omega_0 \in \Omega$ is defined by
\begin{equation}
f'(\omega_0 )={\lim_{h\rightarrow 0}}\frac{f(\omega_0 +h)-f(\omega_0 )}{h}\nonumber
\end{equation}
provided that the limit exists and the domain $\Omega$ is so chosen that $h$ is non-singular in it.
If the bicomplex derivative of $f$ exists at each point of its domain $\Omega$ then $f$ will be a bicomplex holomorphic function in $\Omega$.

Below we list some useful results of different bicomplex valued functions defined on the domain $\Omega$ for $\omega=\omega_1\textbf{e}_1+\omega_2\textbf{e}_2,\varpi=\varpi_1\textbf{e}_1+\varpi_2\textbf{e}_2 \in \Omega$:
\begin{eqnarray}
  &(i)& \omega^n=\omega_1^n \textbf{e}_1+\omega_2^n\textbf{e}_2,\nonumber\\
  &(ii)& e^\omega=e^{\omega_1}\textbf{e}_1+e^{\omega_2}\textbf{e}_2,\nonumber\\
  &(iii)& \cos\omega=\cos\omega_1\textbf{e}_1+\cos\omega_2\textbf{e}_2,\nonumber\\
  &(iv)& \sin\omega=\sin\omega_1\textbf{e}_1+\sin\omega_2\textbf{e}_2,\nonumber\\
  &(v)& \frac{\omega}{\varpi}=\frac{\omega_1}{\varpi_1}\textbf{e}_1+\frac{\omega_2}{\varpi_2}\textbf{e}_2;\quad \varpi \mbox{ is non singular},\nonumber\\
  &(vi)& \omega.\varpi=\omega_1\varpi_1\textbf{e}_1+\omega_2\varpi_2\textbf{e}_2,\nonumber\\
  &(vii)& \int_\Omega f(\omega)d\omega=\int_{\Omega_1}f_1(\omega_1)\textbf{e}_1+\int_{\Omega_2}f_2(\omega_2)\textbf{e}_2,\nonumber\\
  &(viii)& \frac{d}{d\omega}f(\omega)=\frac{d}{d\omega_1}f_1(\omega_1)\textbf{e}_1+\frac{d}{d\omega_2}f_2(\omega_2)\textbf{e}_2.\nonumber
  \end{eqnarray}
  \newpage
  \begin{center}
\textbf{Appendix-B}
\end{center}
  \textbf{Cauchy-Riemann Matrix representation:}\\
If $\omega=x_1+ix_2+\hat{i}x_3+i\hat{i}x_4$ is an element of $\textbf{T}$ in a four-component form we can define a function $\mathcal{N}$  on $\textbf{T}$ as follows:
\begin{equation}\label{f-matr}
\mathcal{N}(\omega)=\left(
                      \begin{array}{cccc}
                        x_1 & -x_2 & -x_3 & x_4 \\
                        x_2 & x_1 & -x_4 & -x_3 \\
                        x_3 & -x_4 & x_1 & -x_2 \\
                        x_4 & x_3 & x_2 & x_1 \\
                      \end{array}
                    \right)
\end{equation}
which is a real Cauchy-Riemann matrix. The set of Cauchy-Riemann matrices with the operations of usual matrix addition and multiplication equipped with the norm $\sqrt{x_1^2+x_2^2+x_3^2+x_4^2}$ is a Banach algebra and is isomorphic and isometric to the bicomplex algebra $\textbf{T}$. Cauchy-Riemann matrices corresponding to the idempotent units $\textbf{e}_1,\textbf{e}_2$ are
\begin{equation}\label{bc-pre-matr}
\varepsilon_1=\left(
                \begin{array}{cccc}
                  \frac{1}{2} & 0 & 0 & \frac{1}{2} \\
                  0 & \frac{1}{2} & -\frac{1}{2} & 0 \\
                  0 & -\frac{1}{2} & \frac{1}{2} & 0 \\
                  \frac{1}{2} & 0 & 0 & \frac{1}{2} \\
                \end{array}
              \right),\quad \varepsilon_2=\left(
                \begin{array}{cccc}
                  \frac{1}{2} & 0 & 0 & -\frac{1}{2} \\
                  0 & \frac{1}{2} & \frac{1}{2} & 0 \\
                  0 & \frac{1}{2} & \frac{1}{2} & 0 \\
                  -\frac{1}{2} & 0 & 0 & \frac{1}{2} \\
                \end{array}
              \right).
\end{equation}
The unique decomposition of $\omega$ in its idempotent representation
\begin{equation}
\omega=\left[(x_1+x_4)+i(x_2-x_3)\right]\textbf{e}_1+\left[(x_1-x_4)+i(x_2+x_3)\right]\textbf{e}_2
\end{equation}
provides the unique decomposition of the corresponding Cauchy-Riemann matrix namely,
\begin{equation}\label{f-matr-decompose}
\mathcal{N}(\omega)=\varepsilon_1 \mathcal{N}\left[(x_1+x_4)+i(x_2-x_3)\right]+\varepsilon_2 \mathcal{N}\left[(x_1-x_4)+i(x_2+x_3)\right]
\end{equation}
Thus we have the forms
\begin{equation}\label{f-matr-dec-1}
\mathcal{N}\left[(x_1+x_4)+i(x_2-x_3)\right]=\left(
                                               \begin{array}{cccc}
                                                 (x_1+x_4) & -(x_2-x_3) & 0 & 0 \\
                                                 (x_2-x_3) & (x_1+x_4) & 0 & 0 \\
                                                 0&0  & (x_1+x_4) & -(x_2-x_3) \\
                                                 0 & 0 & (x_2-x_3) & (x_1+x_4) \\
                                               \end{array}
                                             \right)
\end{equation}
 and
\begin{equation}\label{f-matr-dec-2}
\mathcal{N}\left[(x_1-x_4)+i(x_2+x_3)\right]=\left(
                                               \begin{array}{cccc}
                                                 (x_1-x_4) & -(x_2+x_3) & 0 & 0 \\
                                                 (x_2+x_3) & (x_1-x_4) & 0 & 0 \\
                                                 0&0  & (x_1-x_4) & -(x_2+x_3) \\
                                                 0 & 0 & (x_2+x_3) & (x_1-x_4) \\
                                               \end{array}
                                             \right).
\end{equation}
}

\begin{thebibliography}{99}
\bibitem{2}G.Birkhoff and J.von Neumann \textit{J.Ann.Math.} \textbf{37}, 823 1936.
\bibitem{f1}D.Finkelstein, J.M.Jauch, S.Schiminovich and D.Speiser \textit{J.Math.Phys.} \textbf{3}, 207 1962.
\bibitem{f2}D.Finkelstein, J.M.Jauch, S.Schiminovich and D.Speiser \textit{J.Math.Phys.} \textbf{4}, 788 1963.
\bibitem{AD}S.L.Adler \textit{Quaternion quantum mechanics and quantum fields} Oxford University Press, 1995.
\bibitem{Mar}M.D.Maia \textit{Spin and Isospin in Quaternion Quantum Mechanics} arXiv:hep-th/9904067.
\bibitem{3}C.Segre \textit{Math.Ann} \textbf{40} 467 1892.
\bibitem{4}H.Toyoshima \textit{IEICE Trans.Int.Syst.E} \textbf{80}, 236 1998.
\bibitem{5}I.V.Biktasheva and V.N.Biktasheva, \textit{J.Nonlin.Math.Phys.} \textbf{8}, 28 2001.
\bibitem{6}A.Castaneda and V.V.Kravchenko \textit{J.Phys.A:Math.Gen.} \textbf{38}, 9207 2005.
\bibitem{7}D.Rochon and S.Trembly \textit{Adv.Appl.Clifford Alg.} \textbf{12}, 231 2004.
\bibitem{8}J.Mathieu, L.Marchildon and D.Rochon \textit{Canadian J.Phys.} \textbf{91}, 1093 2013.
\bibitem{ben1}C.M.Bender and S.Boettcher \textit{Phys.Rev.Lett.} \textbf{80} 5243 1998.
\bibitem{ben2}C.M.Bender, D.C.Brody and H.F.Jones \textit{Phys.Rev.Lett.} \textbf{89} 270401 2002.
\bibitem{new}C.M.Bender \textit{Rep.Prog.Phys.} \textbf{70} 947 2007.
\bibitem{sch}L.I.Schiff \textit{Quantum Mechanics} McGraw-Hill, 1968.
\bibitem{n8}C. E. Ruter, K. G. Makris, R. El-Ganainy, D. N.Christodoulides, M. Segev, and D. Kip \textit{Nat. Phys.} \textbf{6}, 192 2010.
\bibitem{n9}A. Szameit, M. C. Rechtsman, O. Bahat-Treidel, and M. Segev \textit{Phys. Rev. A} \textbf{84}, 021806(R) 2011.
\bibitem{n10}A. Regensburger, C. Bersch, M.-A. Miri, G. Onishchukov, D.N.Christodoulides, and U. Peschel \textit{Nature (London)} \textbf{488}, 167 2012.
\bibitem{n11}S. Bittner, B. Dietz, U. G?unther, H. L. Harney, M. Miski-Oglu, A. Richter, and F. Sch\"{a}afer \textit{Phys. Rev. Lett.} \textbf{108}, 024101 2012.
\bibitem{n12}Y. Sun, W. Tan, H. Q. Li, J. Li, and H. Chen \textit{Phys. Rev.Lett.} \textbf{112}, 143903 2014.
\bibitem{n13}B. Peng, S. K. Ozdemir, F. Lei, F. Monifi, M. Gianfreda, G. L.Long, S. Fan, F. Nori, C.M. Bender, and L. Yang \textit{Nat. phys.} \textbf{10},394 2014.
\bibitem{n14}C. M. Bender, M. Gianfreda, S. K. Ozdemir, B. Peng, and L.Yang \textit{Phys. Rev. A} \textbf{88}, 062111 2013.
\bibitem{Kle}S. Kleiman, U. G\"unther  and N. Moiseyev \textit{Phys. Rev.Lett.} \textbf{101}, 080402 2008.
\bibitem{mosr}A.Mostafazadeh \textit{Int.J.Geom.Meth.Mod.Phys.} \textbf{7} 1191 2010.
\bibitem{benr}N.Moiseyev \textit{Non-Hermitian Quantum Mechanics} Cambridge Univ. Press, 2011.
\bibitem{wunner}D.Dast, D.Haag, H.Cartarius, J.Main and G.Wunner \textit{J.Phys.A:Math.Theor.} \textbf{46} 375301 2013.
\bibitem{dast}D.Dast, D.Haag, H.Cartarius, G.Wunner R. Eichler and J.Main  \textit{Fortschr.Phys.} \textbf{61} 124 2013.
\bibitem{cart}H.Cartarius, J.Main and G.Wunner \textit{Phys.Rev.} \textbf{A77} 013618 2008.
\bibitem{wunner2}R. Gut\"ohrlein, J. Main, H.Cartarius and G.Wunner \textit{J.Phys.A:Math.Theor.} \textbf{46} 305001 2013.
\bibitem{wunner1}D.Dizdarevic, D.Dast, D.Haag, J.Main, H.Cartarius and G.Wunner \textit{Cusp bifurcation in the eigenvalue spectrum of $\mathcal{PT}$-symmetric Bose-Einstein condensates} arXiv:1501.03725
\bibitem{kaushal}R S Kaushal \textit{J.Phys.A:Math.Gen.} \textbf{34} L709 2001.
\bibitem{kp}R S Kaushal and Parthasarathi \textit{J.Phys.A:Math.Gen.} \textbf{35} 8743 2002.
\bibitem{price}G B Price \textit{An introduction to multicomplex spaces and functions} Marcel
       Dekkar, 1991.
\bibitem{lav}R.G.Lavoie, L.Marchildon and D.Rochon, \textit{Nuovo Cimento B} \textbf{125}, 1173 2010.
\end{thebibliography}
\end{document}